\documentclass[twocolumn,showpacs,preprintnumbers,amsmath,amssymb,apl,superscriptaddress]{revtex4}
\usepackage{graphicx, subfigure}% Include figure files
\usepackage{dcolumn}% Align table columns on decimal point
\usepackage{bm}% bold math

\newcommand{\bs}{\boldsymbol}

\begin{document}

\title{Double-heterostructure cavities: from theory to design}

\author{Sahand Mahmoodian}
\affiliation{Centre for Ultrahigh bandwidth Devices for Optical Systems (CUDOS), IPOS, School of Physics, University of Sydney, 2006  Australia}
\author{J.E. Sipe}
\affiliation{Department of Physics, University of Toronto, 60 St. George Street, Toronto, ON M5S 1A7, Canada}
\author{Christopher G. Poulton}
\author{Kokou B. Dossou}
\author{Lindsay C. Botten}
\affiliation{CUDOS, School of Mathematical Sciences, University of Technology, Sydney, 2007 Australia}
\author{Ross C. McPhedran}
\author{C. Martijn de Sterke}
\affiliation{Centre for Ultrahigh bandwidth Devices for Optical Systems (CUDOS), IPOS, School of Physics, University of Sydney, 2006  Australia}

\email{sahand@physics.usyd.edu.au}
\date{\today}

\begin{abstract}
We derive a frequency-domain-based approach for radiation (FAR) from double-heterostructure cavity (DHC) modes. We use this to compute the quality factors and radiation patterns of DHC modes. The semi-analytic nature of our method enables us to provide a general relationship between the radiation pattern of the cavity and its geometry. We use this to provide general designs for ultrahigh quality factor DHCs with radiation patterns that are engineered to emit vertically.
\end{abstract}

\pacs{42.70.Qs, 42.25.Bs}
\maketitle

\section{Introduction}
In the 25 years since their conception \cite{john1987strong, yablonovitch1987inhibited}, photonic crystals (PCs) have become an indispensible tool
in modern photonics experiments. One of the principal uses of PCs is in the creation of ultra-high quality factor ($Q$) micro-cavities,
which allow the trapping of light for long periods of time in very small modal volumes ($V)$
\cite{akahane2003high,song2005ultra,kuramochi2006ultrahigh,deotare2009high}. The large ratios of $Q/V$ that can be achieved with PC cavity modes have enabled a range of applications requiring strong light-matter interactions, such as cavity quantum electrodynamics (QED) experiments \cite{yoshie2004vacuum, englund2010resonant, hennessy2007quantum, englund2007controlling}, optical switching \cite{nozaki2010sub}, sensing \cite{loncar2003photonic,kwon2008optimization} and multiple-harmonic generation \cite{galli2010low, mccutcheon2007experimental}.

\begin{figure}[t]
\centering\includegraphics[width=0.45\textwidth]{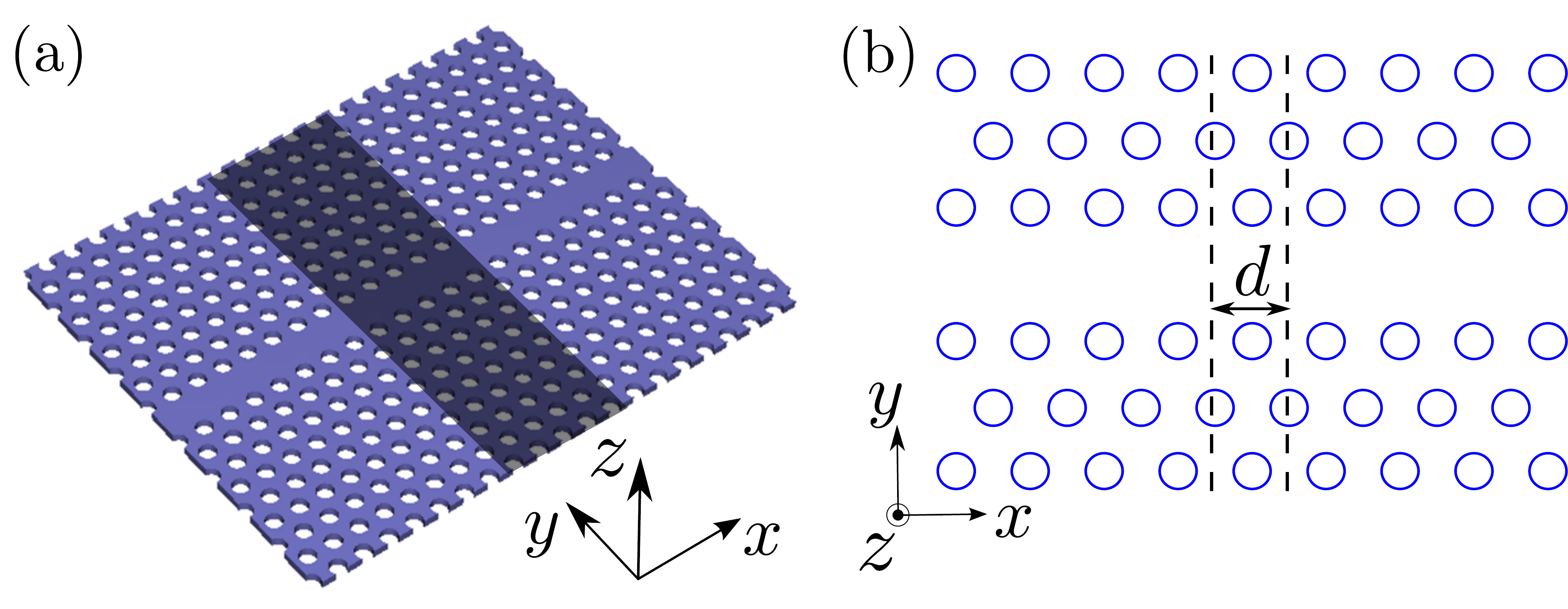}
\caption{\label{PCW_Schematic}(Color online) (a) Schematic of a DHC. In the shaded region the average refractive index is higher than in the remainder of the PCW. (b) Schematic of a finite segment of a PCW that is periodic with period $d$ in the $x$-direction (the finite out-of-plane thickness is not shown). The unit cell is the region between the broken lines and has period $d$.}
\end{figure}

The current state-of-the-art in PC cavities is the Double-Heterostructure Cavity (DHC) \cite{song2005ultra}. DHCs are important because of their ultra-high $Q$ factors and also because the cavities can be readily integrated with other photonic components. DHCs are realized by increasing the average refractive index of a Photonic Crystal Waveguide (PCW) slab in a strip-like region (see Fig. \ref{PCW_Schematic}(a)). This increase often involves a small perturbation, and is achieved by manipulating the geometry of the PCW lattice \cite{song2005ultra, kuramochi2006ultrahigh, kwon2008ultrahigh}, or by increasing the refractive index of either the slab \cite{tomljenovic2007high, lee2009photowritten} or the holes \cite{tomljenovic2006design, smith2007microfluidic, bog2008high}. Within the perturbed region the modes in the PCW are above cutoff and so can propagate, however in the
unperturbed PCW the modes are evanescent, thus leading to trapping of the guided mode. Recent experiments have demonstrated DHC Q-factors of $Q>10^6$ \cite{taguchi2011statistical}.

The development of planar PC cavities was accompanied by a discussion on the exact mechanisms of loss in these structures, and, by implication,
how best to manipulate the $Q$. All planar PC cavity modes are intrinsically lossy because they radiate in the out-of-plane direction \cite{Joannopoulos:2008:PCM:1628775};
in 2003 Noda {\it et al.} introduced the idea of {\em gentle confinement}, in which it was argued that the radiation from the cavity, and hence the loss, could be computed
using the overlap of the Fourier components of the cavity mode with the light-cone. To maximize the $Q$, the
cavity can be constructed to have a mode with a Gaussian-like envelope with a width of several periods
\cite{akahane2003high, song2005ultra, asano2006ultrahigh}, in order to minimize the extent of the cavity mode in Fourier space.
In the context of planar PC cavities this is analogous to the work of Englund {\it et al.} \cite{englund2005general}, and Vu\u{c}kovi\'c {\it et al.} \cite{vuckovic2002optimization}, who showed that the far-field properties of a cavity mode can be computed using the fields above a PC slab.
However, Sauvan {\it et al.} questioned the validity of this approach, arguing that the dominant radiative loss occurred due to
waveguide impedance mismatch at the cavity boundaries, leading to Fabry-Perot reflections that dominate the loss \cite{sauvan2004photonics}.
The understanding of mode confinement and radiation in planar PC cavities has thus-far been hampered by the lack of a comprehensive theory
of confined states in these structures.
%In the work of Englund and and Vu\u{c}kovi\'c, for example, the underlying fields are first computed using Finite Difference Time Domain (FDTD) calculations, which for high $Q$ cavities involves lengthy computations and gives little insight into the underlying physics.

Here, we present a first-principles theory of DHC modes in 3D planar PC structures, an approach that we designate the
Frequency Approach to Radiation (FAR) method. In a recent short publication, we used the FAR to compute the $Q$ factors and radiation patterns of DHC modes \cite{mahmoodian2012first}.
In this paper we provide a detailed description of the FAR and use it to provide simple designs for DHCs with radiation patterns which have been engineered to emit predominantly in the vertical direction.

The FAR consists of two parts: (i) we use a truncated basis of bound PCW modes that lie outside the light cone to construct a non-radiating approximation for the DHC mode. We use a Hamiltonian method for our mode expansion and generalize our previous theory \cite{mahmoodian2012first} by including non-rotating wave terms. (ii) We apply perturbation theory to this non-radiating approximate DHC mode to compute the Fourier components of the polarization field $\mathbf P$ that lie inside the light cone. These are then used to compute the far-field radiation pattern and $Q$ factor \cite{sipe1987new}. This method is efficient as it splits the task of solving a computationally intensive problem into solving smaller problems that are straightforward to compute. Bound PCW modes can be computed rapidly using well-established numerical methods \cite{johnson2001block}, and once these are known, the perturbation theory takes approximately $15$ minutes using our MATLAB code. This method's computational efficiency derives from avoiding the need to compute radiative modes directly. The FAR is semi-analytic and provides insight
into the nature of the radiative processes of planar cavity structures. A central feature of the FAR is an integral equation
that contains a driving term relating the geometry of the DHC to its modes' radiation fields, which contain much of the qualitative features of the radiation pattern. Through an examination of this term, we provide several designs for realizing ultrahigh $Q$ factor DHCs with modes that radiate predominantly in the vertical direction.

This paper is structured as follows: Section \ref{nonRadSection} describes how we expand the cavity modes in a basis of bound PCW modes. The formulation for computing the radiation properties of the DHC is then outlined in Section \ref{radiationSection}, with results presented in Section \ref{results} and discussed further in Section \ref{qualitativeInsight}. Then in Section \ref{designsSection} we use the FAR to design DHCs which radiate predominantly in the vertical. In Section \ref{conclusion} we discuss our results and conclude. Appendix \ref{app:Num} outlines our approach for numerically solving the integral equation which is central to FAR.

\section{Non-radiative approximation} \label{nonRadSection}

In this section we describe how we obtain a non-radiating approximation for cavity modes. We outline how we expand a DHC mode using PCW modes, and then show that this provides a good approximation for the mode profile of the DHC mode.

\subsection{Hamiltonian formulation}\label{hamiltonianSection}

We expand the cavity mode in PCW modes using the Hamiltonian formalism of Sipe and co-authors \cite{pereira2002hamiltonian,chak2007hamiltonian,bhat2006hamiltonian}. This method has proven to be useful in deriving quantum optical versions of linear \cite{chak2007hamiltonian} and nonlinear \cite{pereira2002hamiltonian} coupled mode equations in a systematic way, as well as in devising a quantum optical treatment of dispersion and absorption \cite{bhat2001optical}. Here, we are not interested in quantizing the field, but use the formulation as a vehicle for carrying out the field expansion and determining a first approximation for the cavity mode. A further advantage in our application is that it uses the
divergence-free $\mathbf D$ and $\mathbf B$ fields as the primary fields for the basis functions, so that any superposition is also divergenceless.

We start with the macroscopic Maxwell equations
\begin{equation}
\begin{split}
\label{Maxwell}
\frac{\partial \mathbf D}{\partial t} = \nabla \times \mathbf H&, \hspace{10mm}
\frac{\partial \mathbf B}{\partial t} = -\nabla \times \mathbf E,\\
\nabla \cdot \mathbf D = 0&,
\hspace{10mm}
\nabla \cdot \mathbf B = 0,
\end{split}
\end{equation}
with constitutive relations for non-magnetic dielectric media
\begin{equation}
\begin{split}
\mathbf D(\mathbf r, t) &= \epsilon_0 \mathbf E(\mathbf r, t) + \mathbf P(\mathbf r, t),\\
\mathbf B(\mathbf r, t) &= \mu_0 \mathbf H(\mathbf r, t).
\end{split}
\end{equation}
We first consider the unperturbed structure, i.e. a dispersionless, non-absorbing PCW with an isotropic linear response. Since the DHC geometry involves perturbing the underlying PCW (Fig. \ref{PCW_Schematic}(a)), the PCW modes form a natural basis to expand the DHC modes.  Taking $\mathbf D$ to be one of the primary fields, the polarization field is
\begin{equation}
\mathbf P(\mathbf r,t) = \bar{\Gamma}(\mathbf r)\mathbf D(\mathbf r,t),
\end{equation}
where $\bar{\Gamma}(\mathbf r) = (\bar{\epsilon}(\mathbf r) - 1)/\bar{\epsilon}(\mathbf r)$ and $\bar{\epsilon}(\mathbf r)$ is the permittivity distribution of the PCW. The Hamiltonian representing the energy of the field is
\begin{equation}
\label{generalHamiltonian}
{\cal H} = \frac 1 {2\mu_0}\int d \mathbf r \, \mathbf B(\mathbf r)\cdot \mathbf B(\mathbf r) + \frac 1 {2 \epsilon_0}\int d \mathbf r \frac{\mathbf D(\mathbf r)\cdot \mathbf D(\mathbf r)}{\bar{\epsilon}(\mathbf r)}.
\end{equation}
This is a canonical formulation of electromagnetism and commutators are introduced to produce the appropriate dynamics \cite{chak2007hamiltonian}. The equal time commutation relations are
\begin{equation}
\begin{split}
\label{commutators}
&[D^i(\mathbf r), D^j(\mathbf r')]=[B^i(\mathbf r), B^j(\mathbf r')]=0,\\
&[D^i(\mathbf r), B^j(\mathbf r')]=i \hbar \, \epsilon^{ijk}\frac{\partial}{\partial r^k}[\delta(\mathbf r - \mathbf r')],
\end{split}
\end{equation}
where the superscripts $i$,$j$,$k$ indicate cartesian components, $\epsilon^{ijk}$ is the permutation symbol, and repeated superscripts indicate summation. The dynamics of the fields are given by the Heisenberg equations of motion
\begin{equation}
\begin{split}
\label{heisenberg}
i\hbar \frac{\partial \mathbf D}{\partial t} &= [\mathbf D, {\cal H}],\\
i\hbar \frac{\partial \mathbf B}{\partial t} &= [\mathbf B, {\cal H}].
\end{split}
\end{equation}
Equations~(\ref{generalHamiltonian})-(\ref{heisenberg}) reproduce the Maxwell curl equations in~(\ref{Maxwell}). The divergence equations in (\ref{Maxwell}) act as initial conditions, and if satisfied at some time, the dynamic equations ensure that they are satisfied at all times. Note that in a classical framework the commutators are replaced by Poisson brackets (with appropriate factors of $i\hbar$) and the operators become amplitudes.

We take the PCW to point in the $x$-direction (see Fig.~\ref{PCW_Schematic}(b)), so the permittivity satisfies
\begin{equation}
\bar{\epsilon}(\mathbf r + d\,\hat{\mathbf  x}) = \bar{\epsilon}(\mathbf r),
\end{equation}
where $d$ is the period of the PCW. Solutions to Maxwell's equations are then Bloch modes with band index $m$ and Bloch wavevector $k$. The Bloch modes of the PCW form a complete set and can be used to expand any field
\begin{equation}
\label{blochExpansion}
\mathbf D(\mathbf r, t) = \sum_m \int_{BZ} dk \sqrt{\frac{\hbar \omega_{m,k}}{2}}a_{m,k}\mathbf D_{m,k}(\mathbf r)e^{-i \omega_{m,k} t} + c.c.,
\end{equation}
where the integration is over the Brillouin zone (BZ). This expansion includes all modes: those bound to the PCW, those bound to the slab but not the PCW, as well as the continuum of radiative modes that are not bound to the slab. The Bloch modes take the form
\begin{equation}
\label{blochModes}
\mathbf D_{m,k}(\mathbf r)  = \sqrt{\frac{d}{2\pi}}\mathbf d_{m,k}(\mathbf r)e^{ikx},
\end{equation}
where $\mathbf d_{m,k}(\mathbf r)$ is periodic with period $d$. The magnetic field $\mathbf B$ can be written using similar expressions to (\ref{blochExpansion}) and (\ref{blochModes}). The normalization condition for the $\mathbf D$ field is
\begin{equation}
\label{normalization}
\int d\mathbf r \frac{\mathbf D^*_{m',k'}(\mathbf r)\cdot \mathbf D_{m,k}(\mathbf r)}{\epsilon_0\bar{\epsilon}(\mathbf r)} = \delta_{mm'}\delta(k-k'),
\end{equation}
where the integration is over all space. Using the Poisson summation formula, it can be seen that the field is also normalized over the unit cell
\begin{equation}
\label{normalization2}
\int_{\rm cell} d\mathbf r \frac{\mathbf d^*_{m',k'}(\mathbf r)\cdot \mathbf d_{m,k}(\mathbf r)}{\epsilon_0\bar{\epsilon}(\mathbf r)} = \delta_{m m'}.
\end{equation}
The normalization conditions for the $\mathbf B$ field are the same as in (\ref{normalization}) and (\ref{normalization2}), but with $\epsilon_0\bar{\epsilon}(\mathbf r)$ replaced by $\mu_0$. Using Eqs.~(\ref{blochExpansion})-(\ref{normalization2}) with the commutators in Eq. (\ref{commutators}), it can be shown after some manipulation that the operators $a_{m,k}$ and $a_{m,k}^\dag$ in (\ref{blochExpansion}), where $\dag$ denotes the Hermitian conjugate, satisfy the commutation relations
\begin{equation}
\begin{split}
\label{commutators2}
[a_{m,k}, a_{m',k'}]=0,& \hspace{5mm} [a_{m,k}^\dag, a_{m',k'}^\dag]=0\\
[a_{m,k}, a_{m',k'}^\dag]&=\delta_{mm'}\delta(k-k').
\end{split}
\end{equation}
The Hamiltonian can then alternatively be written as
\begin{equation}
\label{hamiltonianOsci}
{\cal H} = \sum_m \int_{BZ} dk \, \hbar \omega_{m,k}\left(a_{m,k}^\dag a_{m,k}+ \frac 1 2\right),
\end{equation}
which is that of a set of harmonic oscillators.

We include the nRW terms using superpositions of the Bloch modes forming standing waves as basis function, rather than the individual Bloch modes themselves. To avoid double counting we discretize the BZ into an even number of equally spaced points $N$ with an equal number of points on the positive and negative halves. This way, the BZ centre and edge are avoided and the points closest to them are $k=\pm \pi/(Nd)$  and $k=\pm(\pi/d \pm \pi/(Nd))$ respectively. By writing the wave equation for a PCW
\begin{equation}
\begin{split}
\label{wavEqn}
\nabla \times \left[\frac{\nabla \times \mathbf B_{m,k}(\mathbf r)}{\bar{\epsilon}(\mathbf r)} \right] &= \left(\frac{\omega_{m,k}}{c} \right)^2 \mathbf B_{m,k}(\mathbf r),\\
{\textrm{with }}\,\,\,\,\,\mathbf D_{m,k}(\mathbf r) &= \frac{i}{\mu_0 \omega_{m,k}}\nabla \times \mathbf B_{m,k}(\mathbf r),
\end{split}
\end{equation}
we note that both the forward propagating Bloch mode $\mathbf B_{m,k}(\mathbf r)$ and the complex conjugate of the backward propagating Bloch mode $\mathbf B^*_{m,-k}(\mathbf r)$ satisfy Eq. (\ref{wavEqn}) (since $\omega_{m,-k} = \omega_{m,k}$). A complex conjugation of the Bloch mode definition (\ref{blochModes}) is equivalent to taking a backward propagating mode. This means that, apart from an overall phase factor, the two modes $\mathbf B_{m,k}(\mathbf r)$ and $\mathbf B^*_{m,-k}(\mathbf r)$ are equivalent. From the second of Eq.~(\ref{wavEqn}), it is clear that this is also true for $\mathbf D_{m,k}(\mathbf r)$ and we can write
\begin{equation}
\mathbf D_{m,-k}^*(\mathbf r) = e^{i\phi_{m,k}}\mathbf D_{m,k}(\mathbf r).
\end{equation}
Re-expressing the terms in Eq.~(\ref{blochExpansion}) with negative $k$ in terms of complex conjugated Bloch modes with positive $k$ and using $\omega_{-k} = \omega_k$ we obtain
\begin{equation}
\begin{split}
\label{forwardD}
\mathbf D(&\mathbf r) = \sum_m\int_{k>0}\!dk\,\sqrt{\frac{\hbar \omega_{m,k}}{2}}(a_{m,k} + a_{m,-k}^\dag e^{i \phi_{m,k}}) \mathbf D_{m,k}(\mathbf r)\\
&+ \sum_m\int_{k>0}\!dk\,\sqrt{\frac{\hbar \omega_{m,k}}{2}}(a_{m,k}^\dag + a_{m,-k} e^{-i \phi_{m,k}}) \mathbf D_{m,k}^*(\mathbf r).
\end{split}
\end{equation}
The superposition of two equivalent, but counter-propagating waves gives a standing wave which can be made a real function, manifested here by the addition of a function and its complex conjugate. We define the standing wave basis
\begin{equation}
\left[ \begin{array}{c} \mathbf C_{m,k}(\mathbf r) \\ \mathbf S_{m,k}(\mathbf r)\\ \end{array} \right] = \frac 1 2 \begin{bmatrix}  1 & 1 \\ -i  & i \end{bmatrix}
\left[ \begin{array}{c} \mathbf D_{m,k}(\mathbf r) \\ \mathbf D_{m,k}^*(\mathbf r) \end{array} \right],
\end{equation}
where these functions are two orthogonal sine and cosine-like functions. In terms of these modes Eq.~(\ref{forwardD}) becomes
\begin{equation}
\begin{split}
\label{modeExp1}
\mathbf D(\mathbf r) = \sum_{m} \int_{k>0} dk \, \sqrt{2} \, Q_{c,m,k} \mathbf C_{m,k}(\mathbf r)\\
+ \sum_{m}\int_{k>0} dk\,\sqrt{2} \, Q_{s,m,k} \mathbf S_{m,k}(\mathbf r),
\end{split}
\end{equation}
where we define new operators through
\begin{equation}
\left[ \begin{array}{c} Q_{c,m,k} \\ Q_{s,m,k}\\ P_{c,m,k} \\ P_{s,m,k}  \end{array} \right] = \sqrt{\frac \hbar {4 \omega_{m,k}}} \begin{bmatrix}  1& 1 & 1 & 1 \\ i & i & -i & -i \\ -i  & i  & i & -i \\ 1 & -1 & 1 & -1 \end{bmatrix}
\left[ \begin{array}{c} a_{m,k} \\ a_{m,-k}^\dag e^{i \phi_{m,k}} \\ a_{m,k}^\dag \\ a_{m,-k}e^{i \phi_{m,k}} \end{array} \right].
\end{equation}
As the notation suggests, $Q$ and $P$ act like canonical position and momentum operators respectively. The consequences of using modes from only half of the Brillouin zone ($k>0$) is having to use  two sets of independent operators, i.e. those with $c$ subscripts and those with $s$ subscripts. It can be shown from Eq.~(\ref{commutators2}) that these satisfy the commutation relations
\begin{equation}
\begin{split}
[Q_{c,m,k},Q_{s,m',k'}] &= 0,\,\,\,\,\, [P_{c,m,k},P_{s,m',k'}] = 0,\\
[Q_{c,m,k},P_{s,m',k'}] &= 0,\,\,\,\,\, [Q_{s,m,k},P_{c,m',k'}] = 0,\\
[Q_{c,m,k}&,P_{c,m',k'}] = i \hbar \delta_{mm'} \delta_{kk'},\\
[Q_{s,m,k}&,P_{s,m',k'}] = i \hbar \delta_{mm'} \delta_{kk'}.
\end{split}
\end{equation}
From the new operators, potential and kinetic energy-like terms can be defined such that the Hamiltonian in (\ref{hamiltonianOsci}) can be re-written as
\begin{equation}
\label{newHamilt}
{\cal H} = \frac 1 2 \sum_{m} \int_{k>0} dk \sum_p \,(P_{p,m,k}^2 + \omega_{m,k}^2 Q_{p,m,k}^2),
\end{equation}
where $p\in\{c,s\}$. This is equivalent to (\ref{hamiltonianOsci}), but now written in terms of the new operators $P$ and $Q$. In compact form, the mode expansion (\ref{modeExp1}) becomes
\begin{equation}
\label{modeExpansionNEW}
\mathbf D(\mathbf r, t) = \sqrt{2}\sum_\alpha \omega_\alpha Q_\alpha \mathbf F_\alpha(\mathbf r),
\end{equation}
where the index $\alpha=(p,m,k)$, i.e. for brevity we use a sum over $\alpha$ to replace the sum and integral in Eq.~(\ref{modeExp1}), and $\mathbf F_\alpha(\mathbf r)=\mathbf C_{m,k}(\mathbf r)$ if $p=c$ and $\mathbf F_\alpha(\mathbf r)=\mathbf S_{m,k}(\mathbf r)$ if $p=s$.

Thus far we have expressed the modes of the unperturbed PCW using a new notation and we have shown that it is consistent with the Hamiltonian in (\ref{hamiltonianOsci}). We now show that this formulation enables us to construct a Hamiltonian for the DHC while keeping the nRW terms. After introducing the cavity into the PCW the Hamiltonian is written ${\cal H}_{\rm cav} = {\cal H} + V$, where
\begin{equation}
\begin{split}
\label{perurbation}
V &= \frac{1}{2\epsilon_0}\int d\mathbf r \, \left(\frac{1}{\bar{\epsilon}(\mathbf r) + \tilde{\epsilon}(\mathbf r)}  - \frac{1}{\bar{\epsilon}(\mathbf r)} \right)\mathbf D(\mathbf r) \cdot \mathbf D(\mathbf r)\\
&\equiv \int d\mathbf r \,\,\, \gamma(\mathbf r) \,\mathbf D(\mathbf r) \cdot \mathbf D(\mathbf r).
\end{split}
\end{equation}
Substituting the mode expansion (\ref{modeExpansionNEW}) into Eq.~(\ref{perurbation}), the perturbation term takes the form
\begin{equation}
V = 2\sum_{\alpha,\beta} \omega_\alpha \omega_\beta Q_\alpha Q_\beta \int d \mathbf r \, \gamma(\mathbf r)\mathbf F_\alpha (\mathbf r)\cdot \mathbf F_\alpha (\mathbf r),
\end{equation}
In terms of the canonical position and momentum the Hamiltonian is
\begin{equation}
\label{hamiltonianPQ}
{\cal H}_{\rm cav} = \frac 1 2 \sum_\alpha P_\alpha^2 + \frac 1 2\sum_{\alpha,\beta} Q_\alpha L_{\alpha\beta}Q_\beta,
\end{equation}
where
\begin{equation}
\label{eigenEquation}
L_{\alpha \beta} = \omega_\alpha^2 \delta_{\alpha \beta} + 4 \omega_\alpha \omega_\beta \int d \mathbf r \, \gamma(\mathbf r)\mathbf F_\alpha(\mathbf r)\cdot \mathbf F_\beta (\mathbf r).
\end{equation}
The coupling of the basis PCW modes is manifested by the off-diagonal terms in $L_{\alpha \beta}$. We note that the product $Q_\alpha L_{\alpha\beta}Q_\beta$ includes the nRW terms.

\begin{figure}[thpb]
\centering\includegraphics[width=0.45\textwidth]{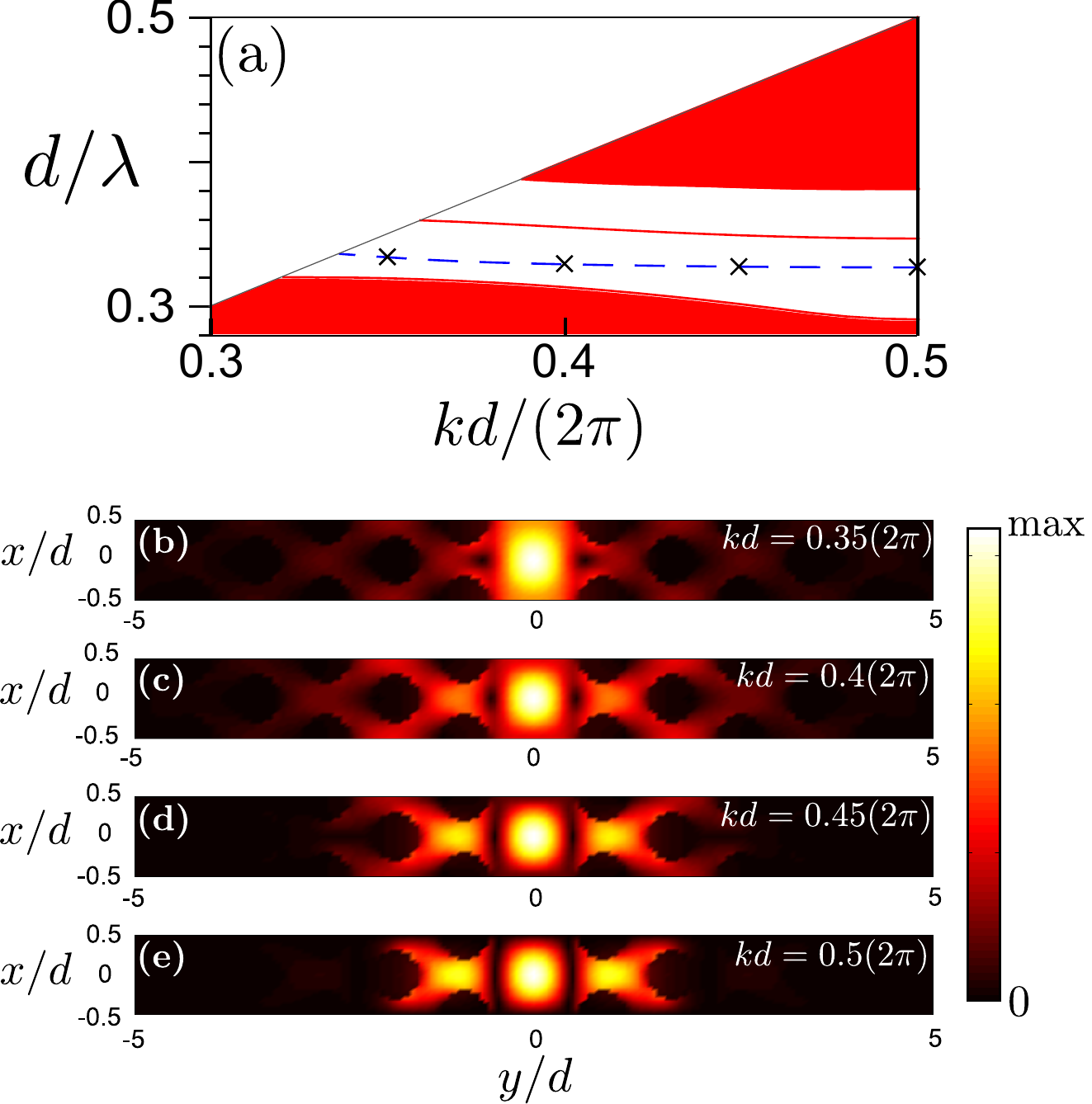}
\caption{\label{PCW_bands} (Color online) (a) TE-like band diagram of the PCW underlying Cavity 2. Broken blue line is the even PCW band used in the mode expansion. Red lines and shading show other modes. (b)-(e) $|\hat{\mathbf y} \cdot \mathbf D_k(\mathbf r)|$ for different bound PCW modes indicated in (a) the crosses. The color scale is linear.}
\end{figure}

Double-heterostructure cavities are formed by slightly perturbing the refractive index of a PCW. Thus DHC modes have frequencies near the edge of the PCW band \cite{song2005ultra}, and the modal fields extend over many period in the direction parallels to the PCW \cite{asano2006ultrahigh}, so their Fourier transform is strongly localized around the BZ-edge. This means that the introduction of the cavity only couples PCW Bloch modes with $k$ values near the BZ-edge. The frequency of DHCs being near the PCW band-edge also means that PCW modes of different bands couple weakly. Thus, to good approximation, we can expand DHC modes using only bound modes from the even PCW band. Such a band is shown in Fig.~\ref{PCW_bands}(a) and the field profiles of some modes in this band are shown in Figs.~\ref{PCW_bands}(b)-(e). These Bloch modes are the basis functions from which we construct the DHC mode, and their field patterns thus control the shape of the cavity mode. As discussed in Sect.~\ref{conclusion}, the variations of these Bloch are important for the cavity mode's far-field properties.

We compute the discrete elements of $L_{\alpha \beta}$ in (\ref{eigenEquation}) numerically. If the BZ is discretized such that of the $N$ points in the BZ, there are $M/2$  positive values of the Bloch wavevector $k>0$ below the light line, the sums over $\alpha$ contain $M$ terms. $L_{\alpha \beta}$ in (\ref{eigenEquation}) is then a real, symmetric $M\!\times\!M$ matrix. Its eigenvalues and eigenvectors correspond, respectively, to the square of the frequency of the cavity modes and the associated real-valued amplitudes of the $\mathbf F_\alpha(\mathbf r)$. Diagonalizing $L$ results in a discrete spectrum of bound cavity modes, as well as a continuum of waveguide states that are not bound to the cavity. We are interested only in the fundamental cavity mode, which corresponds to the lowest eigenvalue.

To diagonalize the eigenvalue equation, we write it as
\begin{equation}
\label{eigenEq1}
\sum_\beta L_{\alpha \beta} s_\beta^{(\gamma)} = \hat{\omega}_\gamma^2 s_\alpha^{(\gamma)},
\end{equation}
and define an orthogonal matrix of eigenvectors
\begin{equation}
\mathrm S = \begin{bmatrix}  s_1^{(1)} & s_1^{(2)} & \ldots & s_1^{(M)} \\ s_2^{(1)} & s_2^{(2)} & \ldots & s_2^{(M)} \\ \vdots  & \vdots  & \ddots & \vdots \\ s_M^{(1)} & s_M^{(2)} & \ldots & s_M^{(M)} \end{bmatrix},
\end{equation}
where $\mathrm S^T \mathrm S = \mathrm S \mathrm S^T = \mathrm I$. We can therefore write
\begin{equation}
\label{matEq}
\mathrm L \mathrm S = \mathrm S \Omega,
\end{equation}
where $\Omega = {\rm diag}(\hat{\omega}_\gamma^2)$ is a diagonal matrix of cavity mode frequencies. Defining new canonical coordinates and momenta $q_\alpha = \mathrm S_{\alpha \beta}^T Q_\beta$ and $p_\alpha = \mathrm S_{\alpha \beta}^T P_\beta$, we obtain a truncated form of the cavity Hamiltonian
\begin{equation}
{\cal H}_{\rm cav} = \frac 1 2 \sum_\alpha (p_\alpha^2 + \hat{\omega}_\alpha^2 q_\alpha^2),
\end{equation}
where the new coordinate and momentum operators satisfy the appropriate commutation relations. Finally, the DHC modes are given by
\begin{equation}
\begin{split}
\label{NRWmode}
\mathbf D(\mathbf r, t) &= \sqrt{2} \sum_{\alpha, \beta} \omega_\alpha \mathbf F_\alpha (\mathbf r) S_{\alpha \beta} \, q_\beta(t)\\
&= \sqrt{2} \sum_{\alpha, \beta} \omega_\alpha \mathbf F_\alpha (\mathbf r) S_{\alpha \beta} \, q_\beta(0) \cos(\hat{\omega}_\beta t).
\end{split}
\end{equation}
This means that the spatial distribution of the fundamental mode $\beta=1$ is
\begin{equation}
\label{NRWmodeFund}
\mathbf D^a(\mathbf r) = \sqrt{2}\sum_\alpha \omega_\alpha \mathbf F_\alpha(\mathbf r) S_{\alpha \, 1}.
\end{equation}
This expression represents of the DHC mode in terms of the bound PCW mode basis with the nRW wave terms retained. Using this representation the total electromagnetic energy in the fundamental cavity mode can be approximated in terms of $L_{\alpha\beta}$ and its eigenvectors as
\begin{equation}
\label{energy}
U \sim \sum_\alpha \omega_\alpha^2 S_{\alpha\, 1}^2 + \sum_{\beta,\alpha} S_{\alpha\,1} S_{\beta\,1}\left(\frac{L_{\alpha \beta} - \omega_\alpha^2\delta_{\beta \alpha}}{2} \right).
\end{equation}
For the cavities we have examined, we have found that the contribution of the second term in (\ref{energy}) is negligible.

\subsection{Mode calculations} \label{modeCalcs}

\begin{figure}[t]
\centering\includegraphics[width=0.5\textwidth]{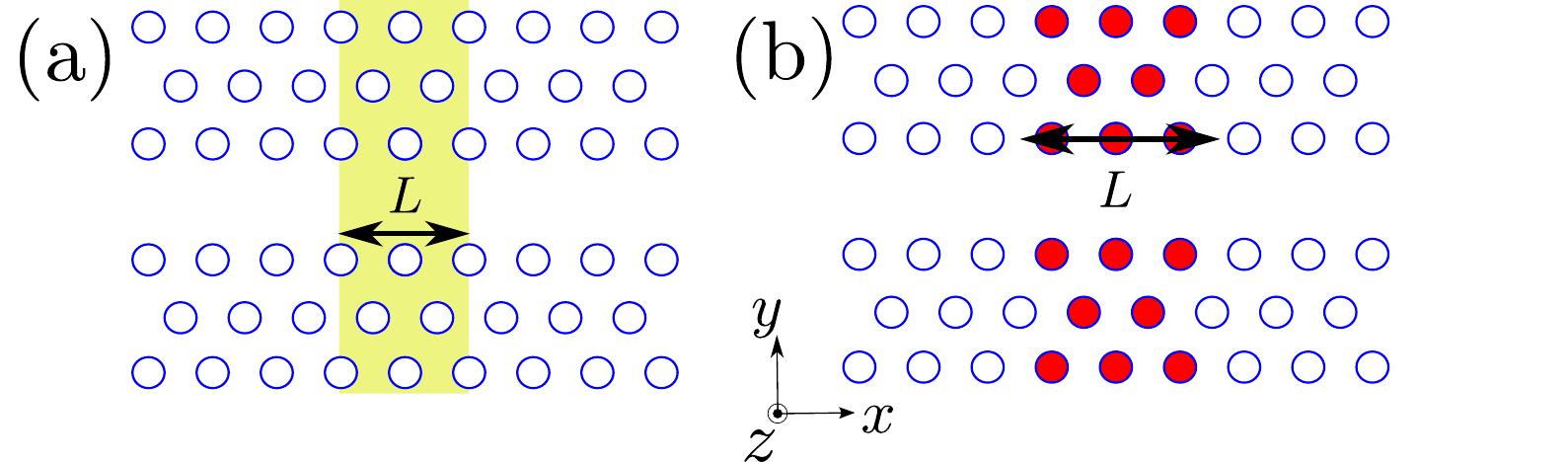}
\caption{\label{DHC_Schematic} (Color online) Schematics of the two different cavity types we consider. (a) Photosensitive cavity: yellow shading represents a local change in the refractive index of the PCW slab creating the cavity. (b) Fluid infiltrated cavity: the red shading indicates a change in refractive index of the holes.}
\end{figure}

In this section, we compare profiles of DHC modes computed using the Hamiltonian method with those computed using FDTD. Double-heterostructure cavities have been realized in different ways \cite{song2005ultra,kuramochi2006ultrahigh,kwon2008ultrahigh,tomljenovic2006design, smith2007microfluidic, bog2008high, tomljenovic2007high,casas2010liquid}. Here we investigate the two geometries shown schematically in Fig. \ref{DHC_Schematic}. In the first (Fig. \ref{DHC_Schematic}(a)), the {\it photosensitive cavity}, the refractive index of the PC slab is uniformly increased, often through a photo-induced refractive index change \cite{tomljenovic2007high,lee2009photowritten}. The second is the {\it fluid infiltrated cavity} where the refractive index of the holes is increased, as can be achieved by fluid infiltration \cite{smith2007microfluidic, bog2008high, casas2010liquid}.

For these two cavity geometries we use three different sets of parameters for our calculations:
\begin{enumerate}
\item A fluid infiltrated cavity based on a $W0.98$ PCW with background index $n_b=3.46$ (consistent with silicon at $\lambda \sim 1500\,{\rm nm}$, where $\lambda$ is the wavelength), air hole radius $a=0.26d$, where $d$ is the period, and slab thickness $t=0.49d$. The cavity is introduced by increasing the refractive index of the holes by $\Delta n_i = 0.2,0.4,0.6$.
\item A photosensitive cavity based on a $W1$ PCW with background index $n_b=2.7$ (the refractive index of some photosensitive chalcogenide glasses \cite{lee2007photosensitive}), air hole radius $a=0.3d$ and slab thickness $t=0.7d$. The cavity is written by uniformly increasing the refractive index of the slab by $\Delta n_p = 0.02,0.04$.
\item A photosensitive cavity based on a $W0.9$ PCW with background index $n_b=3.46$, air hole radius $a=0.3d$ and slab thickness $t=0.7d$. The cavity is introduced by uniformly increasing the refractive index of the slab by $\Delta n_p = 0.02,0.04$.
\end{enumerate}
Cavity~1 is similar to the experimental geometry in \cite{bog2008high} and Cavity 2 is similar to that in \cite{lee2009photowritten}. Geometries such as Cavity~3 have not been realized but could result from ion bombardment of a semiconductor \cite{hines1965radiation, tomljenovic2009flexible}.

\begin{figure*}[t]
\centering\includegraphics[width=\textwidth]{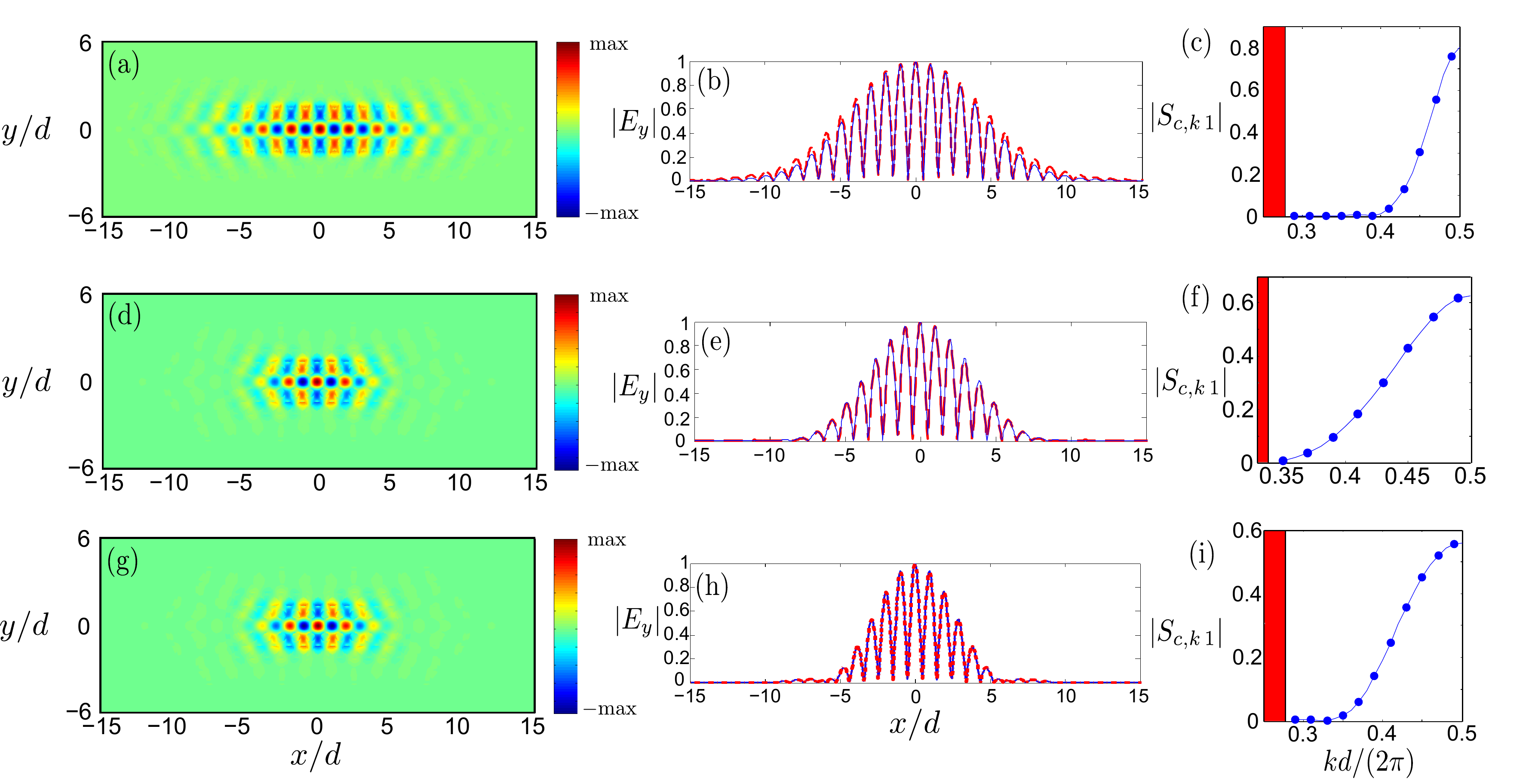}
\caption{\label{DHCmodeProf} Mode profiles of DHC modes taken at $z=0$ and their Bloch mode coefficients. Left column: $E_y$ field computed using the Hamiltonian formulation. Centre column: comparison of $|E_y|$ (taken at $y=z=0$) computed using the Hamiltonian formulation (broken red curve) and using FDTD (blue curve). Right column: magnitude of the Bloch mode coefficients (blue dots) $|S_{c,k\,1}|$ for half of the BZ. The continuous curve is an interpolation of $|S_{c,k\,1}|$ and is included to aid the eye. The red shading shows the value of $k$ where the PCW band is inside the light line. (a)-(c) Cavity 1 with length $L=12d$ and index change $\Delta n_i=0.2$; (d)-(f) Cavity 2 with length $L=10d$ and $\Delta n_p=0.02$; (g)-(i) Cavity 3 with length $L=8d$ and $\Delta n_p=0.02$.}
\end{figure*}

Figure \ref{DHCmodeProf} compares DHC modes computed using our Hamiltonian approach with those computed using FDTD. Figures \ref{DHCmodeProf}(a),(d) and (g) show a $z=0$ slice of the electric field components $E_y$ computed by solving Eq.~(\ref{eigenEq1}). Figures \ref{DHCmodeProf}(b),(e),(h) are similar, but are along a $y=0$ slice. There is good agreement both in the modes' envelope and in the underlying rapid oscillations. Considering the modes' complete 2D cross-section leads to the same conclusion. Figures \ref{DHCmodeProf}(c),(f),(i) show the Bloch mode coefficient $|S_{c,k\,1}|$ for the modes lying below the light line. In all cases the magnitude of the Bloch modes $|S_{c,k\,1}|$ is sufficiently small at the left edges of these plots to justify our truncation or Bloch mode basis to those below the light cone. However, Fig.~\ref{DHCmodeProf}(f) indicates that this approximation may break down if the cavity is made shorter than $L=10d$.  This is because the even PCW modes of slabs with a background index of $n_b=2.7$ have a higher frequency than those with $n_b=3.46$ so there are fewer non-radiative basis functions available for the mode expansion.

Closer inspection of Fig.~\ref{DHCmodeProf}(b) shows a slight discrepancy between the modal widths from our theory and from FDTD. This is surprising since we expect the theory to be well-suited for modes such with a full-width at half maximum of approximately $10d$ and a highly localized Bloch mode composition (Figure \ref{DHCmodeProf}(c)). However, in contrast to photosensitive cavities, fluid infiltrated cavities are formed by large refractive index changes ($\Delta n_i\!=\!0.2$ {\it vs} $\Delta n_p\!=\!0.02$) in regions where the electric fields of the PCW modes are weak. The weakness stems from the dielectric nature of the PCW modes, so they are localised in the slab rather than the holes, and from the absence of holes in the waveguide region where the field is strongest. We believe that our results for the fluid infiltrated cavity could be improved by including more basis modes in the mode expansion, for example modes which are not bound to the PCW, or higher order PCW modes. However, our results are sufficiently accurate for qualitative insight into the modes and their radiation patterns.

We now have a good approximation for the shape of the cavity mode $\mathbf D^a(\mathbf r)$ and its frequency $\hat{\omega}_1$, but the mode does not radiate. In the next section, these quantities are used as ingredients to formulate a perturbative treatment for computing the polarization field of DHC modes inside the light cone. These polarization fields are then used to compute the $Q$ factor and far-field radiation patterns.

\section{Radiation field of DHC modes} \label{radiationSection}

Although $\mathbf D^a(\mathbf r)$ provides a good approximation to the field of the cavity mode, it is non-radiating.  This implies that the Fourier transform of $\mathbf D^a(\mathbf r)$ has no Fourier components within the light cone. On the other hand, we can define a polarization field using $\mathbf D^a(\mathbf r)$,
\begin{equation}
\mathbf P^a(\mathbf r) = \Gamma(\mathbf r)\mathbf D^a(\mathbf r),
\end{equation}
where
\begin{equation}
\begin{split}
\Gamma(\mathbf r) &= \frac{\epsilon(\mathbf r) - 1}{\epsilon(\mathbf r)},\\
\bar{\Gamma}(\mathbf r) &= \frac{\bar{\epsilon}(\mathbf r) - 1}{\bar{\epsilon}(\mathbf r)},
\end{split}
\end{equation}
with $\Gamma(\mathbf r) = \bar{\Gamma}(\mathbf r) + \tilde{\Gamma}(\mathbf r)$, which defines $\tilde{\Gamma}(\mathbf r)$. The polarization field $\mathbf P^a(\mathbf r)$ consists of a part $\bar{\Gamma}(\mathbf r) \mathbf D^a(\mathbf r)$ which does not radiate because $\mathbf D^a(\mathbf r)$ is entirely outside the light cone and $\bar{\Gamma}(\mathbf r)$ has the periodicity of the PCW, and a radiating part $\tilde{\Gamma}(\mathbf r)\mathbf D^a(\mathbf r)$. However, $\tilde{\Gamma}(\mathbf r)$ is not periodic and thus when multiplied by $\mathbf D^a(\mathbf r)$ does have Fourier components inside the light cone.

\subsection{Green tensor for calculating radiation} \label{Rad1}

Before continuing we briefly review a formalism introduced earlier \cite{sipe1987new, saarinen2008green, cote2003simple} for computing radiation fields. All material properties are placed in a polarization term, and the dynamic Maxwell equations with harmonic time dependence read
\begin{equation}
\begin{split}
i \omega \mathbf B(\mathbf r) &= \nabla \times \mathbf E(\mathbf r),\\
-\frac{i\omega}{c^2} \mathbf E(\mathbf r) &= \nabla \times \mathbf B(\mathbf r) + i\omega \mu_0 \mathbf P(\mathbf r).
\end{split}
\end{equation}
Here the polarization field is a source term and we wish to find $\mathbf E(\mathbf r)$ and $\mathbf B(\mathbf r)$ for a given polarization distribution $\mathbf P(\mathbf r)$. This implies that we require a Green tensor that propagates a field emitted by a polarization source oscillating at frequency $\omega$. We adopt an earlier formulation \cite{sipe1987new} that expresses the Green tensor using the variables ($k_x$,$k_y$,$z$). In terms of the cavity mode, this distinguishes between modes that radiate $k_x^2 +k_y^2 \leq k_0^2$, and those that are bound to the slab $k_x^2 +k_y^2 > k_0^2$, where $k_0 = \omega/c$.

\begin{figure}[!t]
\centering\includegraphics[width=0.35\textwidth]{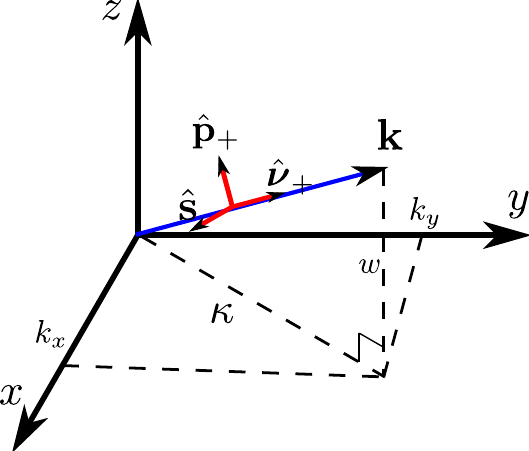}
\caption{\label{spnu_coords} (Color online) Relationship between the cartesian wavevector components $\mathbf k = k_x \hat{\mathbf x} + k_y \hat{\mathbf y} + w \hat{\mathbf z}$ and its $s$ and $p$ polarised components for upward propagating plane waves. For $k_x^2 +k_y^2 > k_0^2$ the diagram is a schematic, since $\hat{\bs \nu}_+$, $\hat{\mathbf  p}_+$, and $\mathbf k$ are all complex.}
\end{figure}

Taking $\boldsymbol \kappa = (k_x, k_y)$ and $\mathbf R = (x,y)$, the relationship between the $\mathbf E(\mathbf r)$ and $\mathbf P(\mathbf r)$ is
\begin{equation}
\label{fieldsFromPolarization}
\mathbf E(\mathbf r) = \int \frac{d \boldsymbol \kappa}{(2 \pi)^2} e^{i\boldsymbol \kappa \cdot \mathbf R} \int dz' G(\boldsymbol \kappa, z-z') \cdot \mathbf P(\boldsymbol \kappa, z'),
\end{equation}
where $\mathbf P(\boldsymbol \kappa, z)$ is the Fourier transform of $\mathbf P(\mathbf r)$ in the $x$ and $y$ variables, and \cite{sipe1987new}
\begin{eqnarray}
\label{greensElectric}
&G(\bs \kappa,z-z';\omega) = \frac{i k_0^2}{2 \epsilon_0 w}\Big( e^{iw(z-z')}\theta(z-z')(\hat{\mathbf s}\hat{\mathbf s} + \hat{\mathbf p}_+\hat{\mathbf p}_+)+\nonumber\\
&e^{-iw(z-z')}\theta(z'-z)(\hat{\mathbf s}\hat{\mathbf s} + \hat{\mathbf p}_-\hat{\mathbf p}_-)\Big) - \frac {1}{\epsilon_0} \delta(z - z') \hat{\mathbf z}\hat{\mathbf z}.
\end{eqnarray}
Here, $\theta(z)$ is the Heaviside-step function, $w = \sqrt{k_0^2 - \kappa^2}$, where $\kappa = |\bs \kappa|$, and $\hat{\mathbf s}$ and $\hat{\mathbf p}_{\pm}$ are the unit vectors for the conventional $s$ and $p$ polarised plane waves respectively, where $+$ and $-$ refer to upward and downward propagation. The definition of $w$ includes the condition that $\sqrt{X}$ is defined so that ${\rm Im}\sqrt{X}\ge0$, and if ${\rm Im}\sqrt{X}=0$, then we take ${\rm Re}\sqrt{X}\ge0$. The directions of $\hat{\mathbf s}$ and $\hat{\mathbf p}_{\pm}$ depend on $\bs \kappa$ and are related to cartesian components by
\begin{equation}
\left[ \begin{array}{c} \hat{\mathbf p}_\pm \\ \hat{\mathbf s} \\ \hat{\bs \nu}_{\pm} \end{array} \right] = \begin{bmatrix}  \frac{\mp k_x}{k_0 \kappa} & \frac{\mp k_y}{k_0 \kappa} & \frac{\kappa}{k_0} \\ \frac{k_y}{\kappa}  & \frac{-k_x}{\kappa} & 0 \\ \frac{k_x}{k_0} & \frac{k_y}{k_0} & \frac{\pm w}{k_0} \end{bmatrix}
\left[ \begin{array}{c} \hat{\mathbf x} \\ \hat{\mathbf y} \\ \hat{\mathbf z} \end{array} \right],
\end{equation}
where $\hat{\bs \nu}_\pm$ points in the direction of propagation. These vectors form the orthogonal  triads $(\hat{\mathbf p}_+, \hat{\mathbf s}, \hat{\bs \nu}_+)$ and $(\hat{\mathbf p}_-, \hat{\mathbf s}, \hat{\bs \nu}_-)$ for upward and downward propagating plane waves respectively. This is shown schematically for upward propagating modes in Fig. \ref{spnu_coords}. The Green tensor contains the outgoing wave condition as required.   The fields for a structure of finite thickness are obtained by convolving the Green tensor with the polarization distribution in the $z$-direction for each value of $\bs \kappa$. The spatial distribution of fields above and below the slab is then obtained by inverse Fourier transform.

Returning to computing the radiation properties, we take the PC slab with thickness $t$ to lie in the region between $-t/2<z<t/2$. Using the Green tensor (\ref{greensElectric}), the electric field above the slab is
\begin{equation}
\label{EandBrad}
\mathbf E_+(\mathbf r) = \int \frac{i d \bs \kappa}{2\pi w}e^{i\bs \nu_+ \mathbf r}\mathbf e_+(\bs \kappa),
\end{equation}
where
\begin{equation}
\mathbf e_+(\bs \kappa) = \hat{\mathbf s}e_+^s(\bs \kappa) + \hat{\mathbf p}_+e_+^p(\bs \kappa),
\end{equation}
and
\begin{equation}
\begin{split}
\label{ep_es}
e_+^s(\bs \kappa) &= \frac{k_0}{4\pi\epsilon_0}\hat{\mathbf s}\cdot \int dz' e^{-iwz'}\mathbf P(\bs \kappa,z')\\
e_+^p(\bs \kappa) &= \frac{k_0}{4\pi\epsilon_0}\hat{\mathbf p}_+ \cdot \int dz' e^{-iwz'}\mathbf P(\bs \kappa,z'),
\end{split}
\end{equation}
with an equivalent expression below the slab. The energy carried away from the cavity at any height above the slab is fixed. Furthermore, an expression for the field at any plane above the slab, say $z=z_0$, can be used to propagate it to any other value of $z>z_0$. This can be used to reproduce the results of Englund {\it et al.} \cite{englund2005general}.

The asymptotic expression for the far-field in spherical polar coordinates is
\begin{equation}
\label{aboveSlab}
\mathbf E(r,\theta,\phi) \sim \mathbf e_+(\bar{\bs \kappa})\frac{e^{i k_0 r}}{r}
\end{equation}
where $\bar{\bs \kappa} = k_0 \, \hat{\mathbf r}\cdot(\hat{\mathbf x}\hat{\mathbf x} +\hat{\mathbf y}\hat{\mathbf y})= k_0(\sin\theta \cos\phi, \sin\theta \sin\phi)$, where $(\theta,\phi)$ are the declination and azimuthal angles respectively. Here, $\hat{\mathbf r}$ is a unit vector that identifies the direction in which we let $r \rightarrow \infty$, and therefore $\bar{\bs \kappa}$ is the projection of $\hat{\mathbf r}$ onto the $xy$ plane multiplied by $k_0$. This then shows that each value of $(k_x,k_y)$ inside the light cone corresponds to radiation in a particular direction $(\theta,\phi)$. The time averaged far-field Poynting vector for an upward travelling wave in spherical coordinates is
\begin{equation}
\label{Poynting}
\langle \mathbf S(r,\theta,\phi, t)\rangle = \frac{2}{\mu_0 c r^2}\left[ |e_+^s(\bar{\bs \kappa})|^2 + |e_+^p(\bar{\bs \kappa})|^2 \right] \hat{\bs \nu}_+,
\end{equation}
where $\langle\,\rangle$ indicates a time average. The quality factor $Q$ is the ratio of energy stored in the cavity mode and the energy lost per cycle and is given by
\begin{equation}
\label{Q}
Q = \omega \frac{U}{2 \int d \phi d\theta \sin\theta \, S(\theta,\phi)},
\end{equation}
where $S(\theta,\phi)= r^2 \langle \mathbf S(r,\theta,\phi, t)\rangle\cdot \hat{\mathbf r}$ is the radial component of the time-averaged Poynting vector as a function of the angles $(\theta,\phi)$ and the integration is over the upper hemisphere. Here, $U$ is the total energy of the field in the DHC mode, approximated by Eq.~(\ref{energy}). Since the geometry is up-down symmetric with respect to the $z$-direction, the factor of $2$ accounts for radiation emitted in the upper and lower hemisphere.

\subsection{The radiative polarization field of DHC modes}\label{INTeqSection}

Computing the far-field properties of DHC modes using the theory presented in Section \ref{Rad1} requires an expression for the Fourier components of the polarization field that lie inside the light cone. However, it would be wrong to approximate $\mathbf P(\mathbf r) = \mathbf P^a(\mathbf r)$ and to substitute its Fourier transform into Eqs.~(\ref{ep_es}). This is clear from $\mathbf P^a(\mathbf r) = \bar{\Gamma}(\mathbf r)\mathbf D^a(\mathbf r) + \tilde{\Gamma}(\mathbf r)\mathbf D^a(\mathbf r)$. We consider the terms $\bar{\Gamma}(\mathbf r)$, $\bar{\epsilon}(\mathbf r)$ and $\mathbf D^a(\mathbf r)$ as being zeroth order, while the terms associated with the perturbation creating the cavity $\tilde{\Gamma}(\mathbf r)$ and $\tilde{\epsilon}(\mathbf r)$ are considered first-order small. Henceforth, all parameters denoted by overbars are zeroth order small, while those with a tilde are first order small. Since $\bar{\Gamma}(\mathbf r)\mathbf D^a(\mathbf r)$ does not contribute to radiation, the radiative components of $\mathbf P^a(\mathbf r)$ are first-order small. Of course $\mathbf D^a(\mathbf r)$ is just an approximation for the actual field $\mathbf D(\mathbf r)$ with corrections terms included by writing
\begin{equation}
\mathbf D(\mathbf r) = \mathbf D^a(\mathbf r) + \mathbf D^c(\mathbf r),
\end{equation}
where $\mathbf D^c(\mathbf r)$ contains first and higher order corrections. The polarization field is then given by
\begin{equation}
\begin{split}
\label{Pcor}
\mathbf P(\mathbf r) &=\mathbf P^a(\mathbf r) + \mathbf P^c(\mathbf r)\\
&= \bar{\Gamma}(\mathbf r)\mathbf D^a(\mathbf r) + \tilde{\Gamma}(\mathbf r)\mathbf D^a(\mathbf r) + \mathbf P^c(\mathbf r),
\end{split}
\end{equation}
where the leading order term $\bar{\Gamma}(\mathbf r)\mathbf D^a(\mathbf r)$ is non-radiating and $\mathbf P^c(\mathbf r)$ contains first-order and higher order corrections. This means that $\mathbf P^c(\mathbf r)$ contains terms that are of the same order as $\tilde{\Gamma}(\mathbf r)\mathbf D^a(\mathbf r)$, and thus an expression for  $\mathbf P^c(\mathbf r)$ is required to compute the radiation fields.

We begin by writing down a simple expression for a polarization field
\begin{equation}
\begin{split}
\label{Pconsistent}
\mathbf P (\mathbf r) &\equiv \epsilon_0(\epsilon(\mathbf r)-1)\mathbf E(\mathbf r)\\
&= \epsilon_0(\epsilon(\mathbf r)-1)\int d \mathbf r' \,G(\mathbf r - \mathbf r';\omega) \cdot  \mathbf P(\mathbf r').
\end{split}
\end{equation}
The Green tensor is the same as that in Eq.~(\ref{greensElectric}), but for brevity, we have expressed it using spatial variables. Since the cavity mode radiates and thus has a complex valued frequency, Eq.~(\ref{Pconsistent}) has no solutions for real valued $\omega$. However, from solving (\ref{eigenEq1}) we have a good approximation for the real part of the resonant frequency $\hat{\omega}_1$, and similarly, $\mathbf P^a(\mathbf r)$ provides a starting point for approximating the polarization field. Using Eq.~(\ref{Pcor}) we write Eq.~(\ref{Pconsistent}) as
\begin{eqnarray}
\label{polEq}
\left[ \mathbf P^a(\mathbf r) + \mathbf P^c(\mathbf r)\right] =& \epsilon_0(\epsilon(\mathbf r) - 1)\int d \mathbf r' \, G(\mathbf r - \mathbf r',\hat{\omega}_1 + \tilde{\omega})\nonumber\\
&\cdot\left[ \mathbf P^a(\mathbf r') + \mathbf P^c(\mathbf r')\right],
\end{eqnarray}
where $\tilde{\omega}$ is the complex first order correction to the frequency. We now take a Taylor expansion of the Green tensor in $\omega$ about $\hat{\omega}_1$
\begin{equation}
\label{greenTaylor}
G(\mathbf r - \mathbf r';\omega)= G(\mathbf r - \mathbf r';\hat{\omega}_1) + \frac{\partial G(\mathbf r - \mathbf r';\hat{\omega}_1)}{\partial \omega}\tilde{\omega} + \ldots.
\end{equation}
To proceed, we write the Green tensor at each value of $k$ as $G(\mathbf r - \mathbf r';\hat{\omega}_1) = G(\mathbf r - \mathbf r';\omega_k) + \tilde{G}_k(\mathbf r - \mathbf r';\hat{\omega}_1)$, i.e. a sum of a Green tensor that propagates out fields with frequency $\omega_k$ and a correction term (which is a function of $k$). The frequency of DHC modes is typically close to the PCW band and thus $(\hat{\omega}_1-\omega_k)/\hat{\omega}_1\ll 1$ for Bloch wavevectors that are below the light line. Therefore $\tilde{G}_k(\mathbf r - \mathbf r';\hat{\omega}_1)$ is considered first-order small. The term
\begin{equation}
\begin{split}
&\epsilon_0(\epsilon(\mathbf r)-1)\int d \mathbf r' G(\mathbf r- \mathbf r';\hat{\omega}_1)\cdot\bar{\Gamma}(\mathbf r')\mathbf D^a(\mathbf r')\\
& = \epsilon_0(\epsilon(\mathbf r)-1)\frac{\mathbf D^a(\mathbf r)}{\epsilon_0 \bar{\epsilon}(\mathbf r)}\\
&+ \epsilon_0(\epsilon(\mathbf r)-1)\sum_\alpha c_\alpha \int d \mathbf r' \tilde{G}_k(\mathbf r- \mathbf r';\hat{\omega}_1)\cdot\bar{\Gamma}(\mathbf r')\mathbf F_\alpha(\mathbf r'),
\end{split}
\end{equation}
using Eq.~(\ref{NRWmodeFund}) and $c_\alpha = \sqrt{2}\,\omega_\alpha S_{\alpha \, 1}$. The first term on the RHS is
\begin{equation}
\begin{split}
\epsilon_0 \left(\epsilon(\mathbf r) - 1\right)\frac{\mathbf D^a(\mathbf r)}{\epsilon_0 \bar{\epsilon}(\mathbf r)}
&= \mathbf P^a(\mathbf r) + \frac{\bar{\Gamma}(\mathbf r)\tilde{\epsilon}(\mathbf r)}{\bar{\epsilon}(\mathbf r)}\mathbf D^a(\mathbf r)\\
&+ \frac{\tilde{\Gamma}(\mathbf r)\tilde{\epsilon}(\mathbf r)}{\bar{\epsilon}(\mathbf r)}\mathbf D^a(\mathbf r),
\end{split}
\end{equation}
and thus we have an equation for $\mathbf P^c(\mathbf r)$
\begin{equation}
\begin{split}
&\mathbf P^c(\mathbf r) = \frac{\bar{\Gamma}(\mathbf r)\tilde{\epsilon}(\mathbf r)}{\bar{\epsilon}(\mathbf r)}\mathbf D^a(\mathbf r) + \frac{\tilde{\Gamma}(\mathbf r)\tilde{\epsilon}(\mathbf r)}{\bar{\epsilon}(\mathbf r)}\mathbf D^a(\mathbf r)\\
&+ \epsilon_0(\epsilon(\mathbf r) - 1)\sum_\alpha \, c_\alpha
\int d \mathbf r' \tilde{G}_k(\mathbf r- \mathbf r';\hat{\omega}_1)
\cdot\bar{\Gamma}(\mathbf r')\mathbf F_\alpha(\mathbf r')\\
&+ \epsilon_0(\epsilon(\mathbf r) - 1)\int d \mathbf r' G(\mathbf r - \mathbf r';\hat{\omega}_1)\cdot\left[\tilde{\Gamma}(\mathbf r')\mathbf D^a(\mathbf r') + \mathbf P^c(\mathbf r') \right]\\
&+ \epsilon_0(\epsilon(\mathbf r) - 1)\,\tilde{\omega}\int d \mathbf r' \frac{\partial G (\mathbf r  - \mathbf r';\hat{\omega}_1)}{\partial \omega}\\
&\cdot\left[\bar{\Gamma}(\mathbf r')\mathbf D^a(\mathbf r') + \tilde{\Gamma}(\mathbf r')\mathbf D^a(\mathbf r') + \mathbf P^c(\mathbf r')\right] + \ldots.
\end{split}
\end{equation}
We only take the terms that are first-order small in the tilde variables $\tilde{\epsilon}(\mathbf r)$, $\tilde{\Gamma}(\mathbf r)$, $\tilde{G}_k(\mathbf r - \mathbf r';\omega)$, $\tilde{\omega}$, and the first order polarization correction $\mathbf P^c(\mathbf r)$, giving an integral equation for the polarization field to first order
\begin{equation}
\begin{split}
\label{polEq3}
&\mathbf P_1^c(\mathbf r) = \frac{\bar{\Gamma}(\mathbf r)\tilde{\epsilon}(\mathbf r)}{\bar{\epsilon}(\mathbf r)}\mathbf D^a(\mathbf r)\\
&+ \epsilon_0(\bar{\epsilon}(\mathbf r) - 1)\sum_\alpha \, c_\alpha \int d \mathbf r' \tilde{G}_k(\mathbf r - \mathbf r';\hat{\omega}_1)\cdot\bar{\Gamma}(\mathbf r')\mathbf F_{\alpha}(\mathbf r')\\
&+ \epsilon_0(\bar{\epsilon}(\mathbf r) - 1)\int d \mathbf r' G(\mathbf r - \mathbf r';\hat{\omega}_1)\cdot\left[\tilde{\Gamma}(\mathbf r')\mathbf D^a(\mathbf r') + \mathbf P_1^c(\mathbf r') \right]\\
&+ \epsilon_0 (\bar{\epsilon}(\mathbf r) - 1)\,\tilde{\omega}\int d \mathbf r' \frac{\partial G(\mathbf r - \mathbf r';\tilde{\omega}_1)}{\partial\omega}\cdot\bar{\Gamma}(\mathbf r')\mathbf D^a(\mathbf r').
\end{split}
\end{equation}

\begin{figure}[t]
\centering\includegraphics[width=0.3\textwidth]{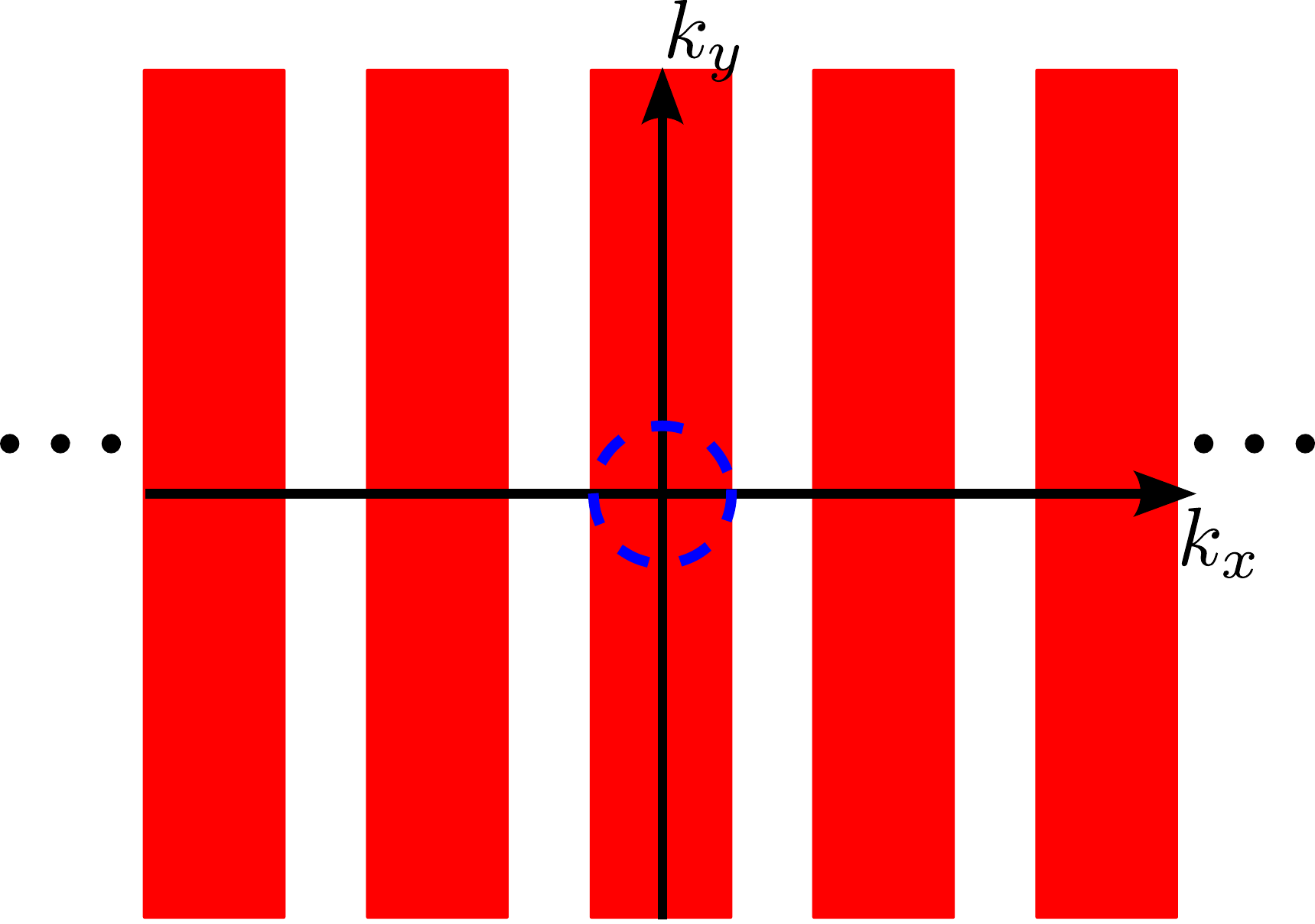}
\caption{\label{Sfilter} (Color online) Schematic of the $\hat{S}$ filter in Fourier space. Red shading indicates the Fourier components that are kept, while white indicates Fourier components that are removed. The light cone is shown at the centre.}
\end{figure}
To compute the radiative polarization field we only require terms with Fourier components inside the light cone, i.e. the first and third terms on the RHS of (\ref{polEq3}). It is tempting to simply apply a filter to both sides of Eq.~(\ref{polEq3}) and eliminate all Fourier components outside the light cone. However, such a filter does not commute with multiplication by $(\bar{\epsilon}(\mathbf r) - 1)$  and therefore can not be taken into the integral. This is because  $\bar{\epsilon}(\mathbf r)$ has the periodicity of the PCW, and so mixes Fourier components separated by $k_x = 2\pi/d$. The requirements of a filter to commute with multiplication by $\bar{\epsilon}(\mathbf r)$ are: it must be periodic in $k_x$ with period  $2\pi/d$, and it must be invariant under translations in $k_y$. We therefore use
\begin{equation}
\label{sFilter}
\hat{S} = \!\int\!\!\frac{d \bs \kappa}{(2\pi)^2}e^{i \bs \kappa \cdot \mathbf R}\!\!\!\sum_{m=-\infty}^{\infty}\!\!\!{\rm rect}\left(\frac{c |k_x|}{2\hat{\omega}_1} + 2 \pi m\right) \int d \mathbf R' e^{-i \bs \kappa \cdot \mathbf R'},
\end{equation}
where ${\rm rect}(x) = 1 $ if $|x| < \frac{1}{2}$ and is zero otherwise. This operates on a function by Fourier transforming, applying a filter in the Fourier domain and then inverse Fourier transforming. The Fourier filter is composed of an infinite series of ${\rm rect}$ functions with width $2\hat{\omega}_1$ separated by $k_x=2\pi/d$ as shown in Fig. \ref{Sfilter}. On application on both sides of Eq.~(\ref{polEq3}), the filter removes all terms without Fourier components inside the light cone. This is because all  terms without Fourier components inside the light cone also do not possess Fourier components separated from the light cone by $k_x=2\pi m/d$, where $m$ is an integer. Equation (\ref{polEq3}) then becomes
\begin{equation}
\begin{split}
\label{polEq4}
&\hat{S}\mathbf P_1^c(\mathbf r) = \hat{S} \frac{\bar{\Gamma}(\mathbf r)\tilde{\epsilon}(\mathbf r)}{\bar{\epsilon}(\mathbf r)}\mathbf D^a(\mathbf r) + \\
& \epsilon_0(\bar{\epsilon}(\mathbf r) - 1)\!\int\! d \mathbf r' G(\mathbf r- \mathbf r';\hat{\omega}_1)\! \cdot\! \hat{S} \left[ \tilde{\Gamma}(\mathbf r')\mathbf D^a(\mathbf r
') + \mathbf P_1^c(\mathbf r')\right].
\end{split}
\end{equation}
By defining $\mathbf P_1^{c,{\rm rad}}(\mathbf r) = \hat{S}\mathbf P_1^c(\mathbf r)$, as well as $\mathbf P_1^{\rm rad}(\mathbf r) = \mathbf P_1^{c,{\rm rad}}(\mathbf r) + \hat{S} \,\tilde{\Gamma}(\mathbf r)\mathbf D^a(\mathbf r)$, Eq.~(\ref{polEq4}) becomes
\begin{equation}
\begin{split}
\label{integralEquation}
&\mathbf P_1^{\rm rad}(\mathbf r) = \hat{S}\left(\left[\frac{\bar{\Gamma}(\mathbf r)\tilde{\epsilon}(\mathbf r)}{\bar{\epsilon}(\mathbf r)}\mathbf + \tilde{\Gamma}(\mathbf r)\right] \mathbf D^a(\mathbf r)\right)\\
&+ \epsilon_0(\bar{\epsilon}(\mathbf r) - 1)\int d \mathbf r' \, G(\mathbf r- \mathbf r';\hat{\omega}_1) \cdot \mathbf P_1^{\rm rad}(\mathbf r').
\end{split}
\end{equation}
This is a Fredholm equation of the second kind whose solution for $\mathbf P_1^{\rm rad}(\mathbf r)$ gives a first order approximation for the radiative components of the DHC mode. The inhomogeneous term evaluates to what degree the perturbation terms $\tilde\epsilon(\mathbf r)$ and $\tilde\Gamma(\mathbf r)$ couple Fourier components from the non-radiative approximation of the field $\mathbf D^a(\mathbf r)$ into the light cone. Once Eq.~(\ref{integralEquation}) is solved for $\mathbf P_1^{\rm rad}(\mathbf r)$, the radiation can be obtained by computing the Poynting vector from Eqs.~(\ref{EandBrad})-(\ref{Q}). Our approach for obtaining solutions to Eq.~(\ref{integralEquation}) is outlined in Appendix \ref{app:Num}.

\section{Radiation calculation results}\label{results}

In this section we present a comparison between $Q$ factors and far-field radiation patterns obtained using the FAR and those computed using fully numerical FDTD calculations.

\begin{figure}[ht]
\centering\includegraphics[width=0.35\textwidth]{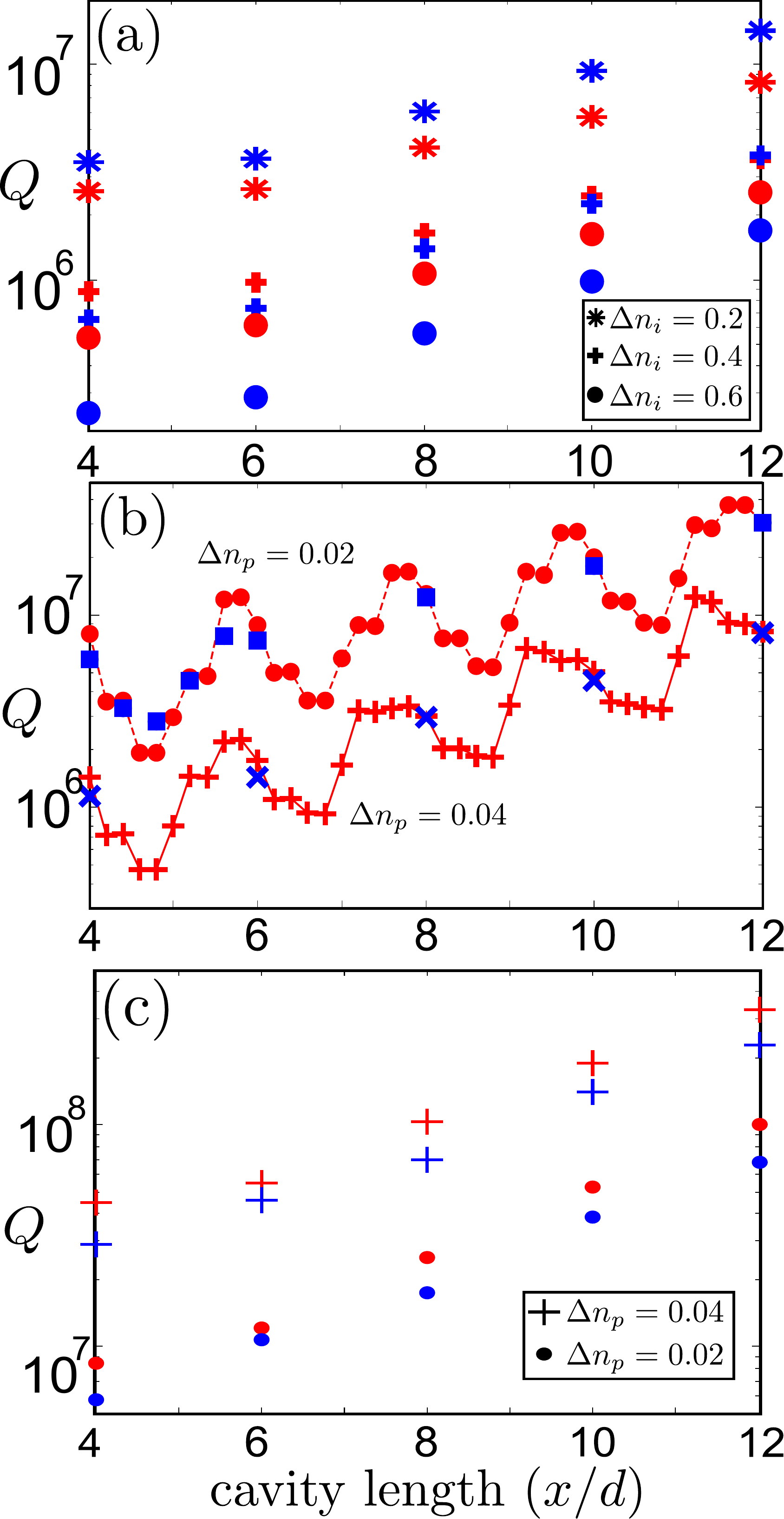}
\caption{\label{Qtriple} (Color online) Quality factors versus cavity length $L$ computed using our FAR method (red symbols) and those computed using FDTD (blue symbols). (a) $Q$ factors for Cavity 1 with $\Delta n_i=0.2,0.4,0.6$. (b) $Q$ factors for Cavity 2 with $\Delta n_i=0.02,0.04$. (b) $Q$ factors for Cavity 3 with $\Delta n_i=0.02,0.04$.}
\end{figure}

We carried out our FDTD calculations using the commercial package $Q$-{\it Finder} (RSoft). The computational parameters were tailored to compute modes with different $Q$ factors, but in all cases we used symmetry to reduce the computation domain to $1/8$ of the DHC. Our computation domain ranged from $(x,y,z) = (0,0,0)$ to $(x,y,z) = (25.5d, \frac{13 d \sqrt{3}}{2}, 1.35d)$, where $d$ is the period. The spatial discretization ranged from $(\Delta x, \Delta y, \Delta z) = d/22$ to $(\Delta x, \Delta y, \Delta z) = d/26$, with adjustments to manage the computation time. The temporal discretization was always set to $c \Delta t=\Delta x/2$. Depending on the parameters, these calculations took between $10$-$50$ hours on a 32 core cluster for each data point in Fig. \ref{Qtriple}.

In Fig.~\ref{Qtriple} we compare the $Q$ factor for the three cavities versus cavity length $L$ computed using the FAR and FDTD. Once the basis functions for the underlying PCW have been computed, the $Q$ factor calculations using the FAR typically takes less than 15 minutes per data point. For all three cavities we obtain good agreement in the trends of the $Q$ factors versus cavity length between FAR and FDTD. For the theoretically calculated $Q$ factors in Figure \ref{Qtriple}(b), we varied the length of the cavity in a continuous manner, while we have only provided values computed using FDTD at a number of intervening points as it is impractical to compute all points using FDTD. Note the large oscillations in $Q$ as the cavity length is varied. We return to this in Sect.~\ref{qualitativeInsight}.

\begin{figure}[ht]
\centering\includegraphics[width=0.4\textwidth]{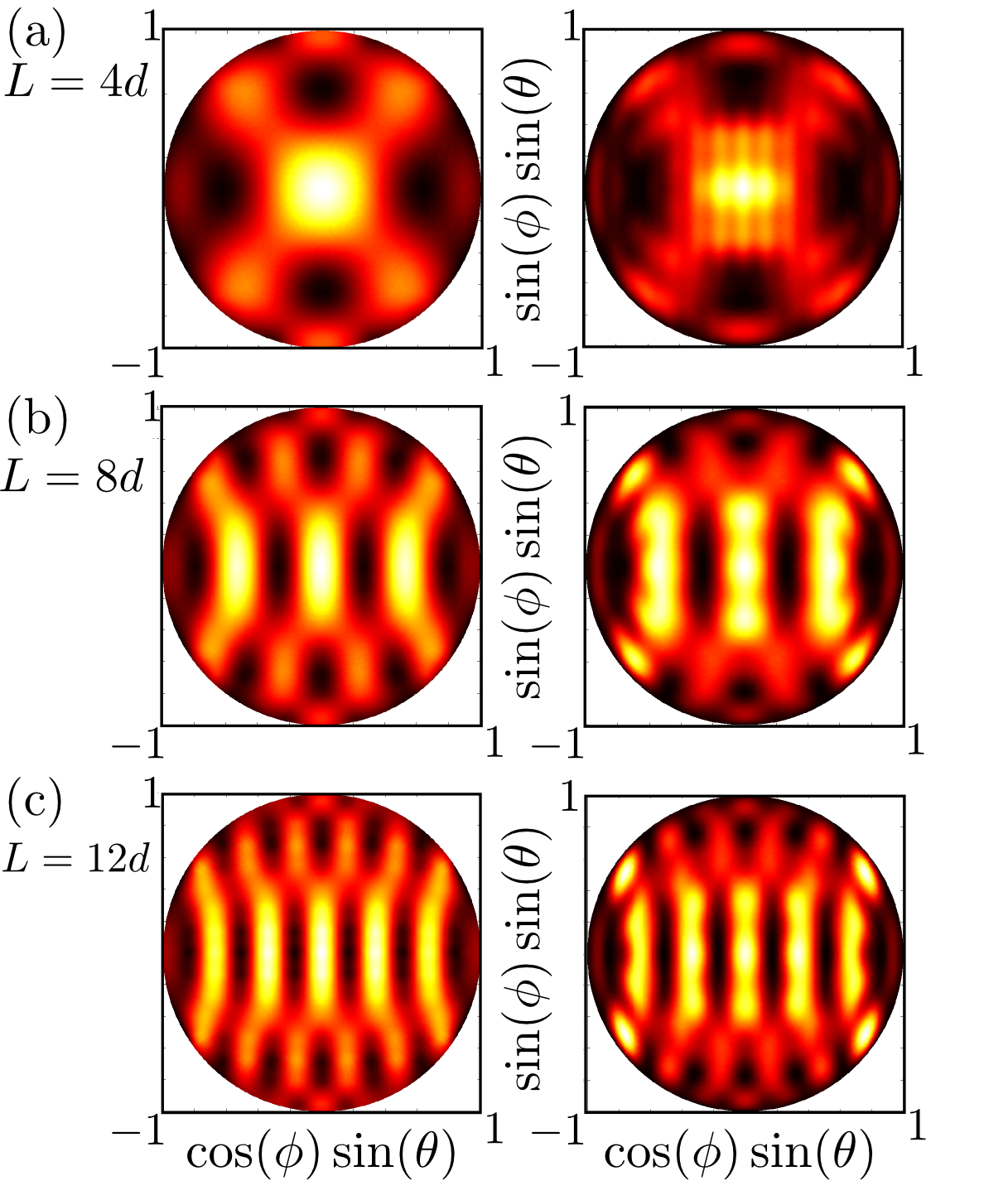}
\caption{\label{SrFluids}(Color online) Far-field Poynting vectors ($S_r$) for Cavity 1 with $\Delta n_i = 0.2$ computed using the FAR (left column) and those computed using FDTD (right column). (a) Far-field radiation pattern for cavity length $L=4d$, (b) $L=8d$ and (c) $L=12d$. Colors as in Figure \ref{PCW_bands}. Here $\phi$ and $\theta$ are the azimuthal and declination angles respectively.}
\end{figure}

In Figure \ref{Qtriple}(b), the difference between $Q$ factors computed using FAR and FDTD is at most $30\%$ ($\sim 2\%$ for their logarithms), while in Figure \ref{Qtriple}(c) the discrepancy is at most $35\%$ ($\sim 2.5\%$ for their logarithms). On the other hand the agreement between FAR and FDTD for Cavity 1 (Figure \ref{Qtriple}(a)) is not as impressive, and the discrepancy is at most a factor of $2$ ($6\%$ for their logarithms). This is likely to be due to the fact that the field profiles computed using the bound mode basis have a slight discrepancy when compared with FDTD (see Figure \ref{DHCmodeProf}(b)). Since the $Q$ factors here are very large, a small discrepancy in $\mathbf D^a(\mathbf r)$ may cause a large change in the radiation properties. Nevertheless, the chief aim of our semi-analytic approach is to obtain qualitative information regarding the general trends in $Q$ factor as a function of cavity parameters, and the results shown in Figure {\ref{Qtriple}} indicate that the theory has achieved this goal.

\begin{figure}[ht]
\centering\includegraphics[width=0.4\textwidth]{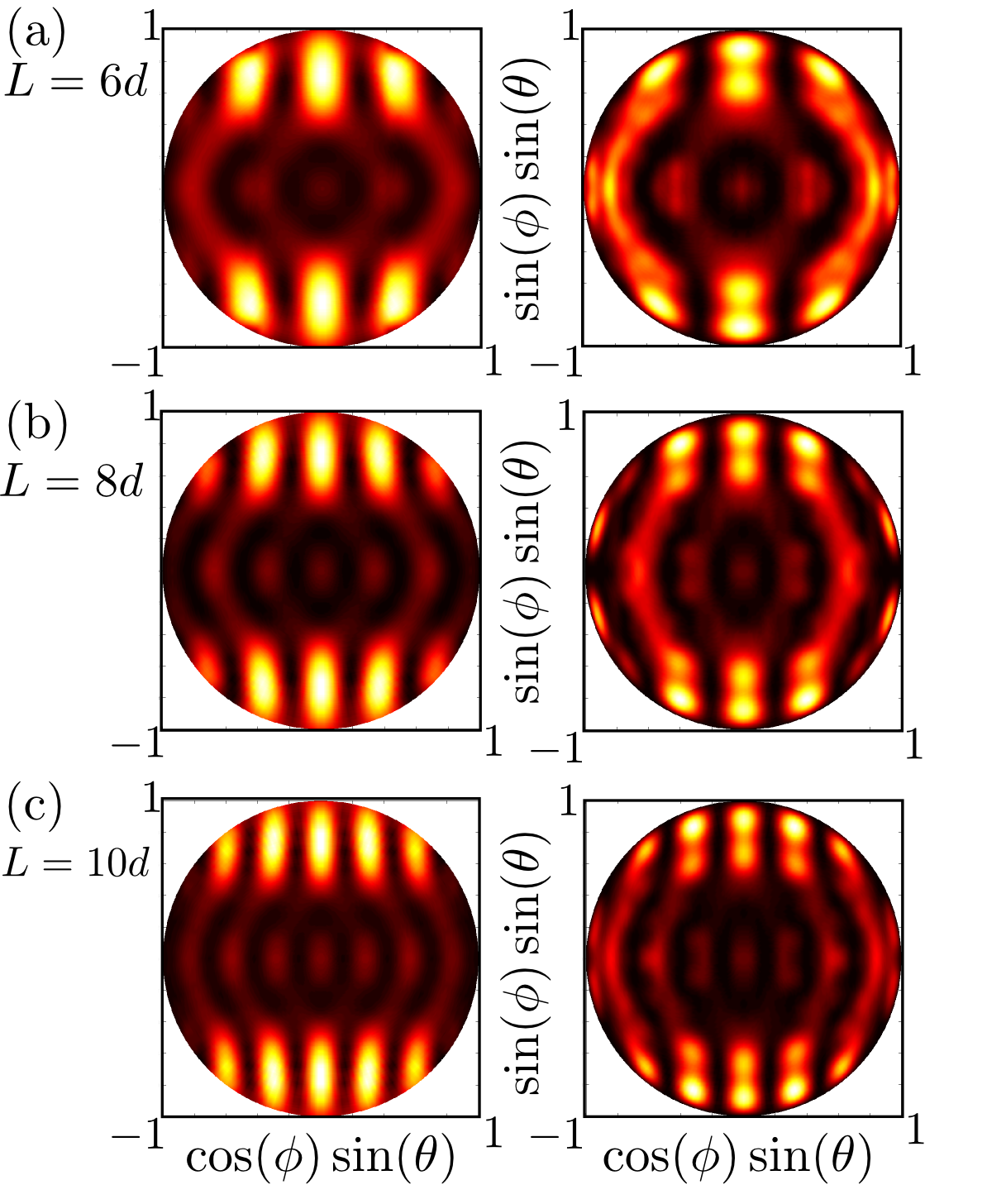}
\caption{\label{SrChalc}(Color online) Far-field Poynting vectors ($S_r$) for Cavity 2 with $\Delta n_p = 0.02$ computed using FAR (left column) and those computed using FDTD (right column). (a) Far-field radiation pattern for cavity length $L=6d$, (b) $L=8d$ and (c) $L=10d$. Colors as in Figure \ref{PCW_bands}.}
\end{figure}
We now examine the far-field radiation patterns (the radial component of the Poynting vector $S_r$) for DHC modes. Figure \ref{SrFluids} shows a comparison of far-field radiation patterns for Cavity 1 computed using the FAR (left column) and those using FDTD (right column). There is good agreement between the two sets of far-field patterns; both show that the number of lobes in the radiation pattern increases as the cavity becomes longer. The cavity with length $L=4d$ has particularly strong radiation in the vertical direction ($\theta \sim 0$), which has been recently shown to be useful for exciting cavity modes from free space \cite{portalupi2010planar}. In Section \ref{designsSection} we provide designs for DHCs whose modes are engineered to emit vertically.

\begin{figure}[htp]
\centering\includegraphics[width=0.4\textwidth]{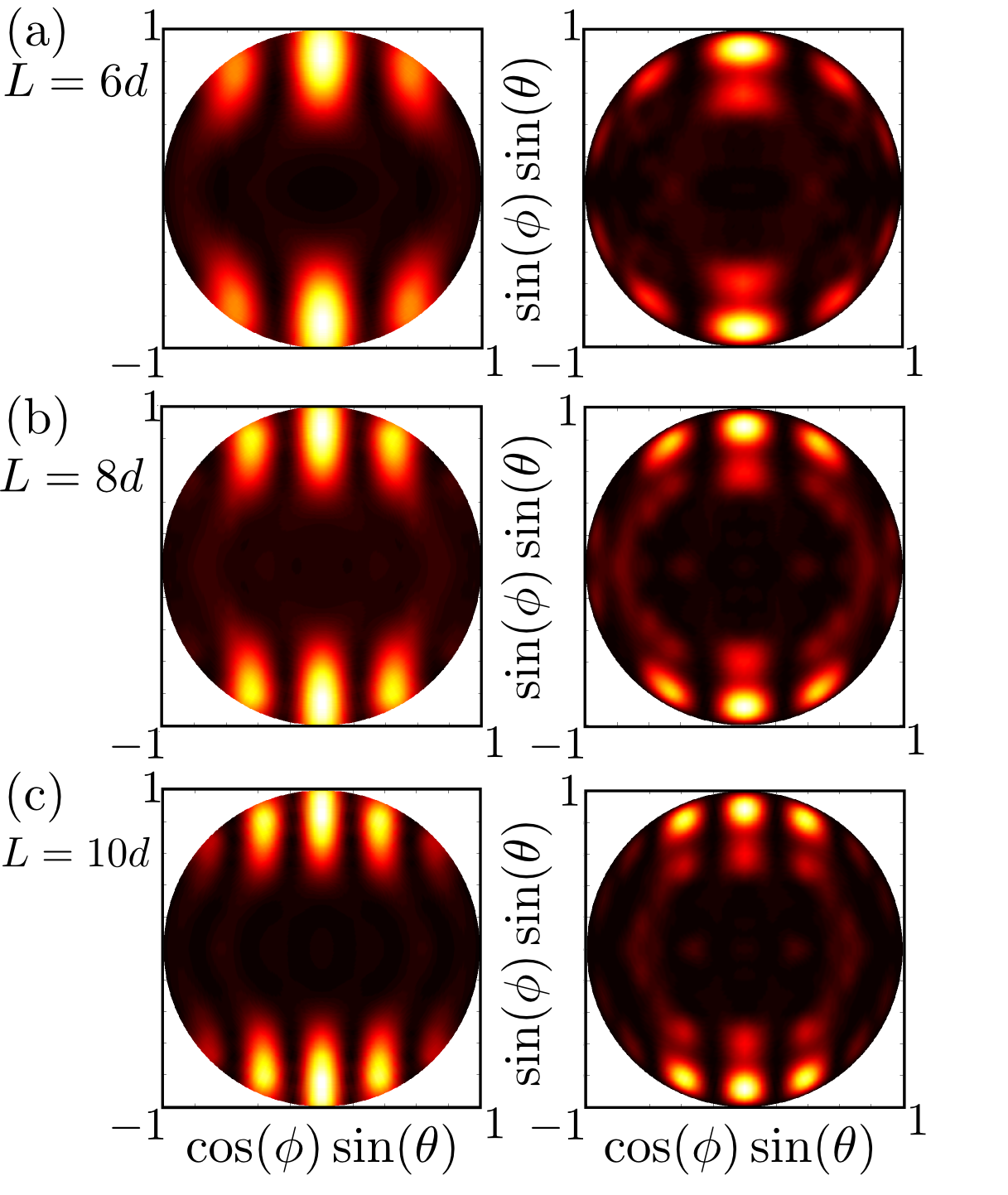}
\caption{\label{SrW9Si}(Color online) Far-field Poynting vectors ($S_r$) for Cavity 3 with $\Delta n_p = 0.02$ computed using the FAR (left column) and those computed using FDTD (right column). (a) Far-field radiation pattern for cavity length $L=6d$, (b) $L=8d$ and (c) $L=10d$. Colors as in Figure \ref{PCW_bands}.}
\end{figure}

Considering now the photosensitive cavity, the far-field radiation patterns of the modes of Cavity 2 are shown in Figure \ref{SrChalc}. Unlike the fluid infiltrated cavity, here the radiation pattern is predominantly directed towards large declination angles $\theta$. The agreement between theory and FDTD is again good as both predict similar radiation directions and both provide the same trends for the number of lobes in the radiation pattern as the cavity length increases. The $Q$ factors of Cavity 2 are larger than those in Cavity 1, even though the refractive index of Cavity 1 is larger ($n_b=3.46$) than that of Cavity 2 ($n_b=2.7$) and the modes of Cavity 1 have envelope functions that vary more slowly than those of Cavity 2 (see Figure \ref{DHCmodeProf}). This is because changes in the background index couple the light much more weakly to the the light cone than changes in the refractive index of the holes. This is clear from examining the Fourier components in the driving term in Eq.~(\ref{integralEquation}). We discuss this in more detail in Section \ref{qualitativeInsight}.

Finally, Figure \ref{SrW9Si} shows the far-field radiation patterns computed for Cavity 3. Again there is good qualitative agreement between theoretical (left column) and numerical results (right column). The radiation patterns here resemble those computed for Cavity 2 in Figure \ref{SrChalc}, however there are fewer lobes in the radiation pattern. This is because the frequency of the modes of Cavity 3 are lower than those for Cavity 2,  and therefore the light cone occupies a smaller region in Fourier space. Again, this can be observed through an examination of the Fourier components of the driving term in Eq.~(\ref{integralEquation}) within the light cone. We discuss this in the following Section.

\section{Analysis of the driving term}\label{qualitativeInsight}

Having established the quantitative capabilities of our theory, we now demonstrate the physical insight available due to its analytic nature. In Eq.~(\ref{integralEquation}), the parameters associated with the cavity geometry $\mathbf D^a(\mathbf r)$, $\tilde{\epsilon}(\mathbf r)$ and $\tilde{\Gamma}(\mathbf r)$, reside in the inhomogeneous driving term $\left[\frac{\bar{\Gamma}(\mathbf r)\tilde{\epsilon}(\mathbf r)}{\bar{\epsilon}(\mathbf r)}\mathbf + \tilde{\Gamma}(\mathbf r)\right]\mathbf D^a(\mathbf r) \equiv \tilde{A}(\mathbf r)\mathbf D^a(\mathbf r)$, while the Green tensor ensures a self consistent interaction between dipoles. We now show that the far-field of the DHC modes can be understood from the Fourier components of the $\tilde{A}(\mathbf r)\mathbf D^a(\mathbf r)$ term below the light line.

\begin{figure}[tp]
\centering\includegraphics[width=0.5\textwidth]{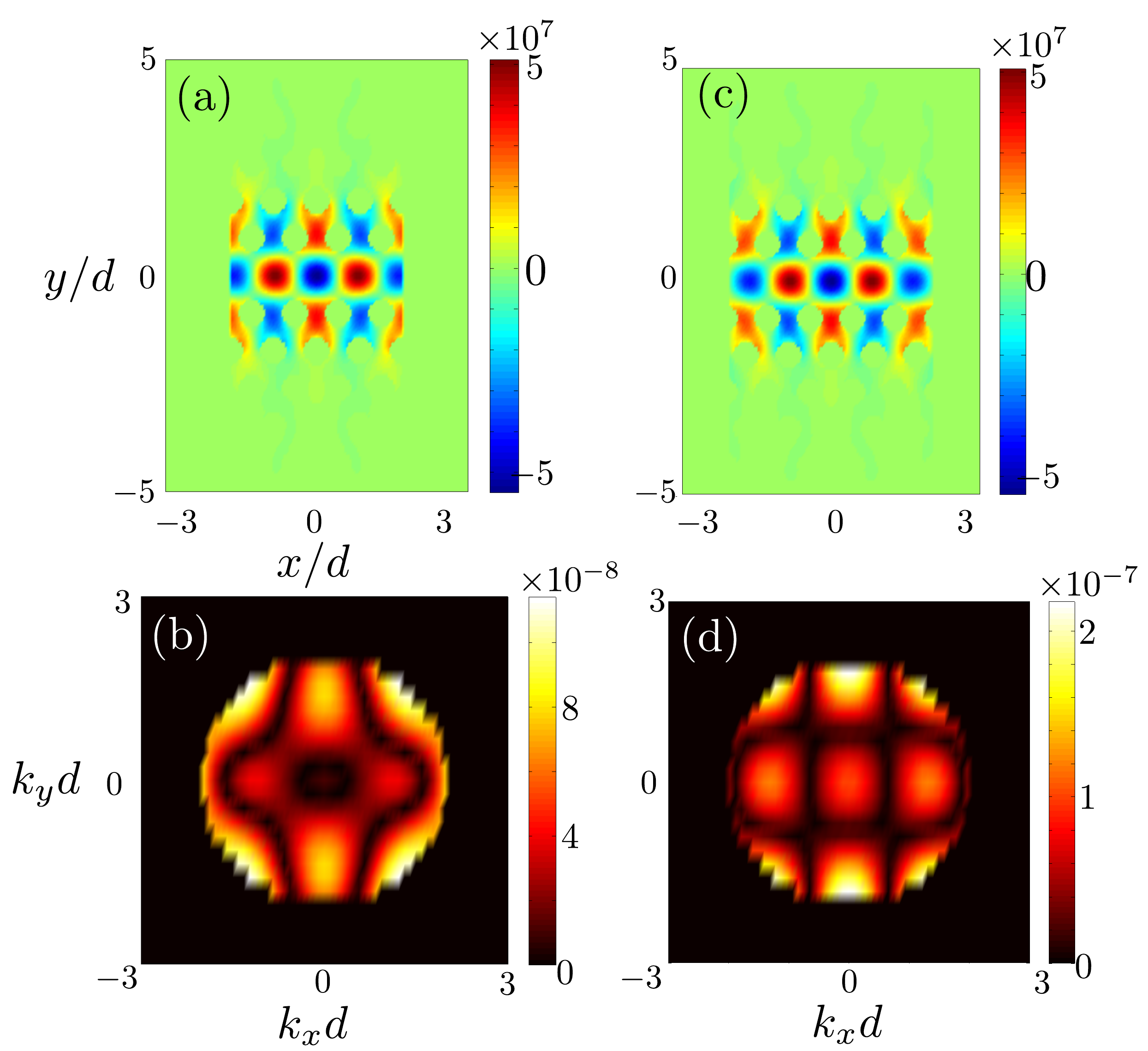}
\caption{\label{GammaTildeQjump}(Color online) Contour plot of the $y$ component of the inhomogeneous term ($z=0$ slice) normalised by the total electromagnetic energy of the mode for Cavity 2 with $\Delta n_p = 0.02$. (a) Cavity length $L=4d$. (b) Cavity length $L=4.8d$. (c) and (d) show the light cone components of the Fourier transforms of (a) and (b) respectively.}
\end{figure}

We first examine the $Q$ factor oscillations Fig.~\ref{Qtriple}(b) associated with sub-period changes in cavity length. Apparently, the magnitude of the radiation is strongly affected by how the cavity perturbation cuts across the PCW modes. To analyze this we normalise the inhomogeneous term $\tilde{A}(\mathbf r)\mathbf D^a(\mathbf r)$ by dividing it by the square root of the total energy in the cavity mode ${\sqrt U}$. The Fourier components of the term $\tilde{A}(\mathbf r)\mathbf D^a(\mathbf r)/{\sqrt U}$ inside the light cone then indicate the strength of the radiating polarization field with respect to the total energy in the cavity mode. Figure \ref{GammaTildeQjump}(a) shows a $z=0$ slice of the $y$ component of $\tilde{A}(\mathbf r)\mathbf D^a(\mathbf r)/{\sqrt U}$ for a cavity with refractive index change $\Delta n_p = 0.02$ and length $L=4d$. Figure \ref{GammaTildeQjump}(c) is similar, but for a longer cavity of length $L=4.8d$. While the results seem similar, the Fourier components that are inside the light cone, shown in Figures \ref{GammaTildeQjump}(b) and (d), differ strongly: the magnitude of the field inside the light cone is considerably larger for the longer cavity,  and we therefore expect that this cavity radiates more strongly than the shorter cavity. This is confirmed in Fig.~\ref{Qtriple}(b), which shows that the $Q$ factor of the cavity with length $L=4d$ is approximately $8$ times larger than that with length $L=4.8d$.

\begin{figure}[ht]
\centering\includegraphics[width=0.5\textwidth]{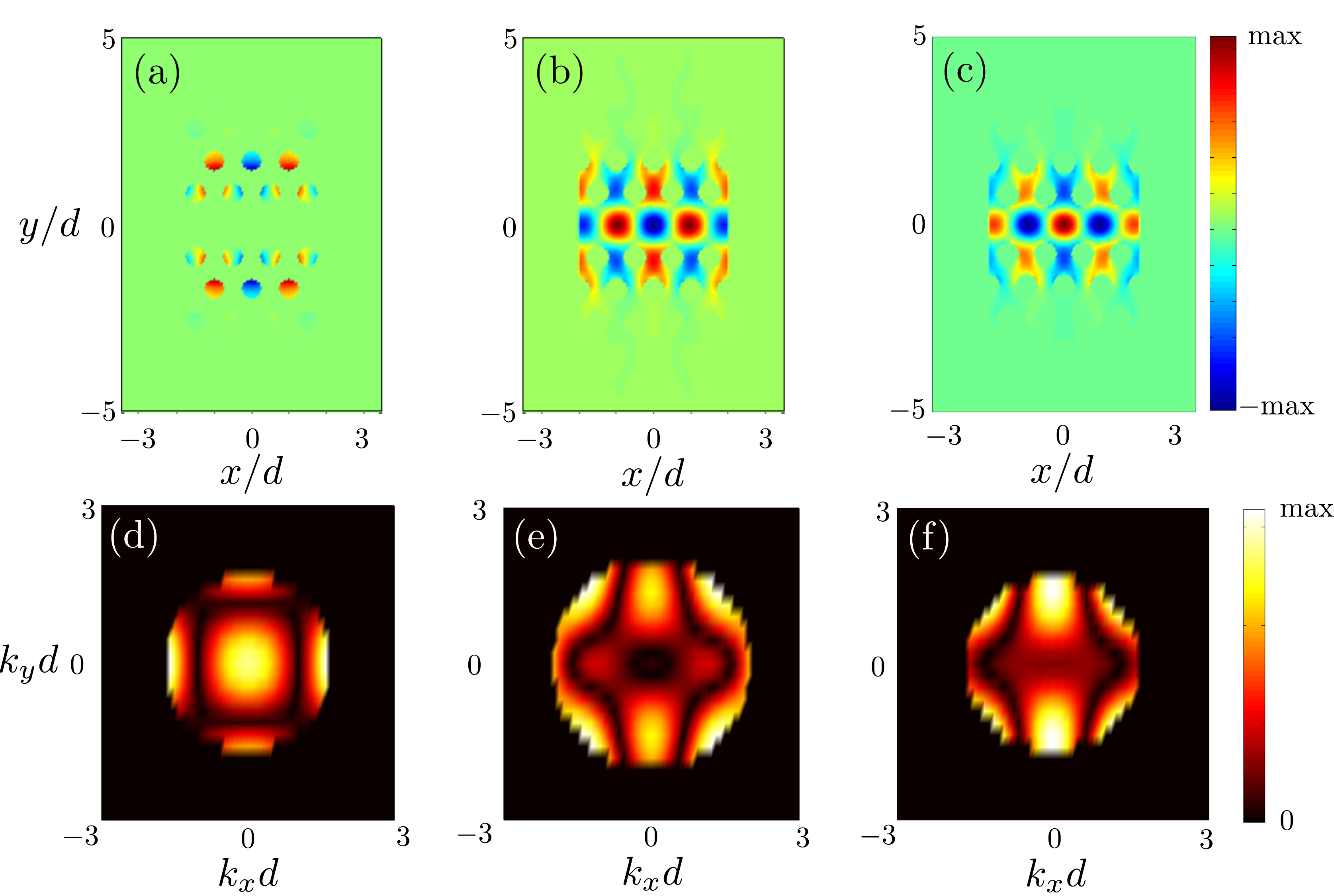}
\caption{\label{GammaTildeDyTriple}(Color online) Contour plot of the $y$ component of the inhomogeneous term ($z=0$ slice)  for (a) Cavity 1 with $\Delta n_i=0.2$ and $L=4d$, (b) Cavity 2 with $\Delta n_p=0.02$ and $L=4d$, and (c) Cavity 3 with $\Delta n_p=0.02$ and $L=4d$ (d),(e),(f) show the light cone components of the Fourier transforms of (a),(b),(c) respectively. Color bars are linear scales.}
\end{figure}

Figures \ref{GammaTildeDyTriple}(a)-(c) show $z=0$ slices of the $\hat{\mathbf y}\cdot \mathbf D^a(\mathbf r)$ component of the inhomogeneous term for the three cavity types. The driving term for Cavity 1 shown in Figure \ref{GammaTildeDyTriple}(a) has a Fourier transform (Figure \ref{GammaTildeDyTriple}(d)) with a strong DC component, indicating strong radiation in the vertical direction consistent with the computed Poynting vector (Figure \ref{SrFluids}(a)). The strong vertical radiation means that this cavity design typically has a smaller $Q$ factor than the photosensitive design. The driving term for Cavity 2 has a Fourier transform that is strongly peaked at the edges of the light cone (Fig.~\ref{GammaTildeDyTriple}(e)) and therefore the radiation is strongest at large declination angles, consistent with Fig.~\ref{SrChalc}. There is also a subtle difference between the far-fields for the two different photosensitive cavities, i.e. for Cavities 2 and 3. The latter have a higher background index implying that its modes have lower frequencies, and consequently the light cone is smaller. This means that fewer features of the driving term overlap the light cone, explaining why Cavity 3 has far-field radiation patterns with fewer lobes (Figure \ref{SrW9Si}) than Cavity 2 (Figure \ref{SrChalc}).

\section{Vertical emission} \label{designsSection}

Recent interest in engineering the radiation pattern of PC cavities \cite{PhysRevB.82.075120, portalupi2010planar, englund2010resonant} has focused on cavities which emit radiation predominantly in the vertical. These not only enable the collection of light exiting the cavity, but allow the cavity mode to be excited from free-space. These cavities were recently used in cavity QED \cite{englund2010resonant, englund2012} and harmonic generation experiments \cite{galli2010low}. We show here that using the FAR we easily arrive at such designs for DHCs.

\begin{figure}[ht]
\centering\includegraphics[width=0.45\textwidth]{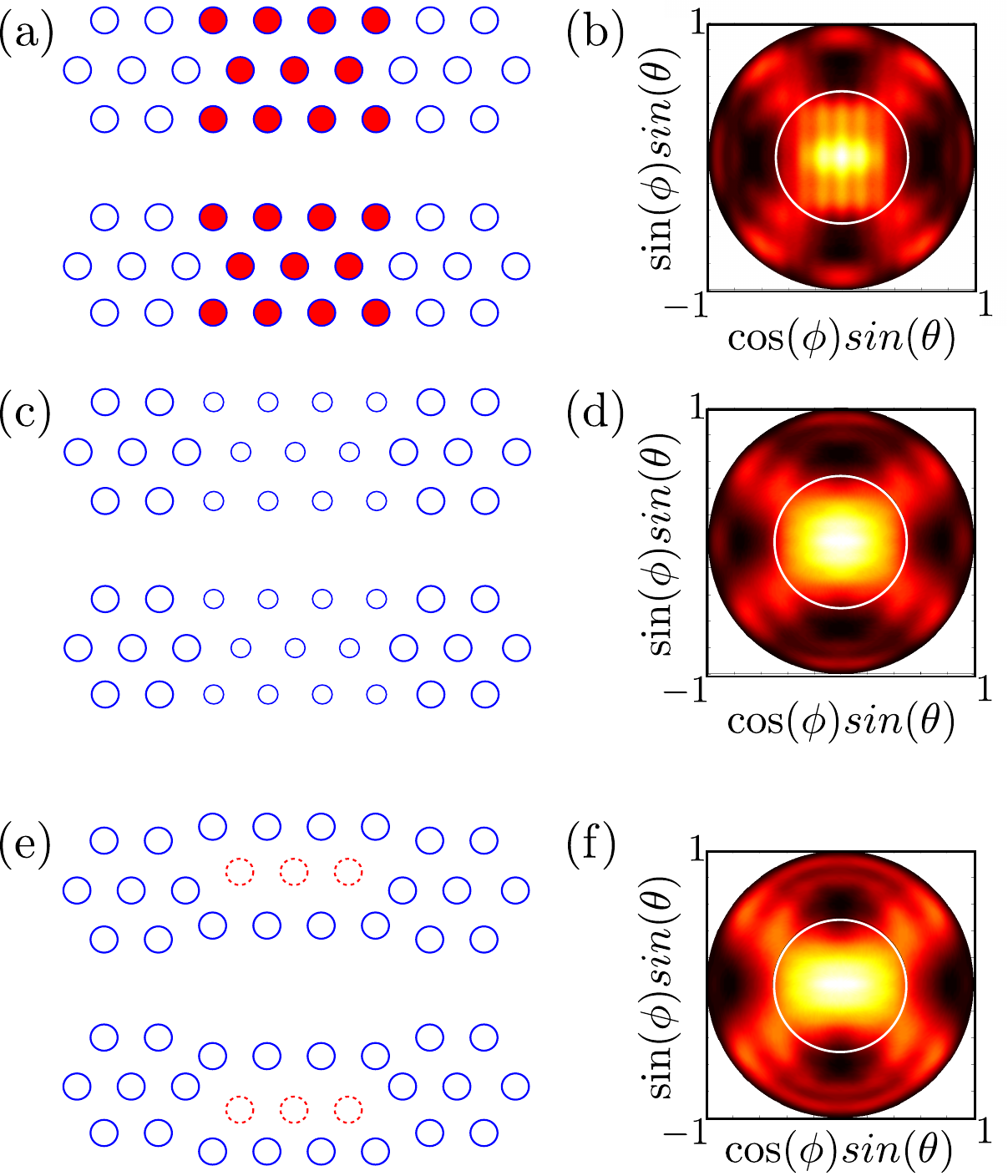}
\caption{\label{designs} (Color online) Schematics of DHC designs that maximize vertical emission and their corresponding radiation patterns ($S_r$). (a)-(b) Fluid infiltrated cavity. (c)-(d) Cavity created by radius change. (e)-(f) Cavity created by hole shift. The holes drawn with dashed red lines are shifted more than others. The white circles in the radiation pattern corresponds to a declination angle of $30^\circ$. The holes shifts and radius changes in (c) and (e) are not to scale. Colors as in Figure \ref{PCW_bands}.}
\end{figure}

As discussed in Sect.~\ref{qualitativeInsight}, the fundamental mode of a DHC to radiate predominantly in the vertical direction if the refractive index is such that $\tilde{A}(\mathbf r)\mathbf D^a(\mathbf r)$ has a non-zero DC Fourier component. While we showed that this is so in a fluid-infiltrated cavity with $L=4d$, this idea applies more generally. Vertical radiation can be achieved by manipulating the holes in a different way. Figure \ref{designs}(a) shows a schematic of a fluid infiltrated cavity of length $4d$ with its radiation pattern computed using FDTD in Fig. \ref{designs}(b). A schematic of a design where the hole radius is decreased is shown in Fig. \ref{designs}(c); the associated radiation pattern in Fig.~\ref{designs}(d), for a structure in which the hole radius was decreased from $0.27d$ to $0.25d$ in a $W0.94$ silicon waveguide with thickness $t=0.45d$, confirms the predominantly vertical emission. We also computed the radiation pattern for a $W0.94$ PCW with holes radius $0.27d$ and thickness $t=0.45d$, where the holes shown in Fig. \ref{designs}(e) are shifted to that of a $W0.98$ PCW and the holes drawn with the dashed red lines are shifted a further $0.02\sqrt{3}d$, i.e. to where the equivalent holes of a $W1.02$ PCW would be. The radiation pattern in Fig. \ref{designs}(f). All three designs have more than $70\%$ of their radiated power within a declination angle of $30^\circ$ (white circles in Figs. \ref{designs}(b),(d),(f)). The computed $Q$ factors for the parameters in Figs \ref{designs}(d),(f) ranged between $2\times10^5$ and $4\times10^5$. Unlike previous designs based on $L3$ cavities \cite{portalupi2010planar}, the $Q$ factor of these designs can be controlled independently of the radiation pattern. The theoretical $Q$ factor of these cavities can be increased by reducing the strength of the perturbation that creates the cavity, i.e. by decreasing the change in radius or hole shift. These cavity designs may have improve the performance of cavity-based experiments in harmonic generation and cavity QED.

\section{Discussion and Conclusion} \label{conclusion}

Previously \cite{mahmoodian2010paired, asano2006ultrahigh}, DHC modes have been characterized by an envelope function that modulates a rapidly varying Bloch mode. Here, we show that this picture is insufficient for explaining some of the physics underlying the results presented here, and that a multiple Bloch mode approach is required.

\begin{figure}[ht]
\centering\includegraphics[width=0.4\textwidth]{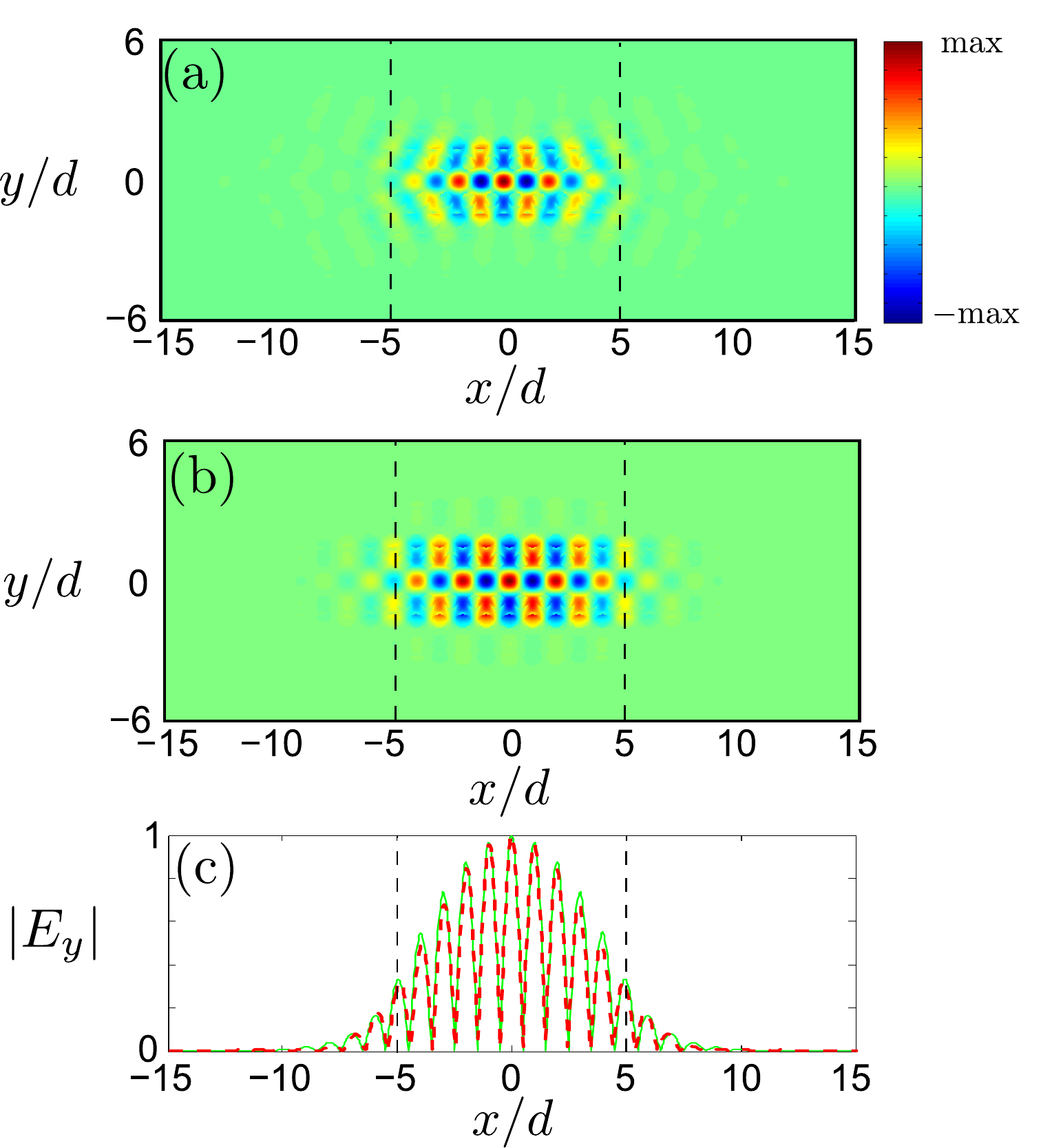}
\caption{\label{SchrodingerVsHamilton} (Color online) Electric field $E_y(\mathbf r)$ for Cavity 2 with length $L=10d$ and $\Delta n_p = 0.02$ computed using (a) the Hamiltonian formulation and (b) the envelope function-based theory. (c)  A $y=z=0$ slice of the magnitude of the $y$-component of the electric field $|E_y|$ computed using the Hamiltonian formulation (dashed red curve) and the envelope-based theory (green curve). The dashed lines show the physical length of the cavity}
\end{figure}

We first compare the modal fields produced using an envelope function formulation with the fields obtained by solving Eq.~(\ref{eigenEq1}). The key difference between the work here and the envelope function picture is that here we construct the cavity mode by superposing multiple Bloch modes, while in the envelope function approach the $\mathbf D$ field of the DHC mode has the form
\begin{equation}
\label{envTheory}
\mathbf D_{\rm env}(\mathbf r) = f(x) \mathbf D_{\pi/d}(\mathbf r),
\end{equation}
where $f(x)$ is the envelope function and $\mathbf D_{\pi/d}(\mathbf r)$ is the displacement field of the band-edge Bloch mode. Figure \ref{SchrodingerVsHamilton}(a) shows the electric field at a $z=0$ slice calculated for Cavity 2 with $L=10d$ and $\Delta n_p = 0.02$ using our Hamiltonian formulation, while Figure \ref{SchrodingerVsHamilton}(b) shows this mode computed using the envelope function-based theory. Even though these modes have almost the same characteristic length, as shown in Figure \ref{SchrodingerVsHamilton}(c), there is a clear difference. The mode computed using the Hamiltonian has a chevron-like feature which is absent from the envelope-based calculation. This chevron-like feature arises because of the difference between the Bloch modes in the mode superposition in Eq.~(\ref{NRWmodeFund}). The fields of the different PCW Bloch modes are illustrated in Figures \ref{PCW_bands}(b)-(e) for the PCW underlying Cavity 2. The important feature here is that the Bloch modes at different Bloch wavevectors have different functional forms with respect to the $y$ variable. This leads to the chevron-like feature of the field profile when these modes superposed to construct the cavity mode through Eq.~(\ref{modeExpansionNEW}). This is not accounted for in the single Bloch mode theory since the functional form of the DHC mode in Eq.~(\ref{envTheory}) only contains a single Bloch mode $\mathbf D_{\pi/d}(\mathbf r)$ modulated by an envelope function $f(x)$ that is only a function of $x$. The $y$-dependence of the cavity mode therefore only lies the band-edge Bloch mode $\mathbf D_{\pi/d}(\mathbf r)$.

In the envelope function theory, changes in the length of the cavity only manifest themselves through the envelope function $f(x)$. Consequently, if $\mathbf D_{\rm env}(\mathbf r)$ is used to compute the Fourier components of  $\tilde{A}(\mathbf r)\mathbf D_{\rm env}(\mathbf r)/{\sqrt U}$ in the light cone, changes in the cavity parameters only affect the $k_x$ distribution and therefore increases in cavity length would lead to a monotonic increase in the $Q$ factor. This is because the characteristic length of $f(x)$ increases as the cavity becomes longer, and hence the Fourier transform of $f(x)$ becomes narrower. The envelope function theory cannot predict the differences in the Fourier components in Figs \ref{GammaTildeQjump}(b) and (d) as these functions differ in both $k_x$ and $k_y$. The $Q$ oscillations are inherently caused by the interference of the different Bloch components of $\mathbf D^a(\mathbf r)$ in the Fourier components of the product $\tilde{A}(\mathbf r)\mathbf D^a(\mathbf r)/{\sqrt U}$ inside the light cone.

The oscillations in $Q$, occurring as a result of superposition of multiple Bloch modes, have a natural interpretation in terms of Fabry-Perot resonances in the waveguide cavity:
as the length of the cavity increases the waveguide impedance at the end `facets' (the plane where the perturbation ends) changes
periodically. This leads to a change in the spacing of the fringes in the $x$-direction of Fourier space, which in turn leads to a periodic modulation of the $Q$ factor.
This supports the interpretation of Sauvan {\it et al.} \cite{sauvan2004photonics}, in which a dominant contribution to the loss of a DHC cavity arises from
Fabry-Perot reflections at the cavity boundaries.

In conclusion, we have presented the first-principles FAR for calculating the near-field and far-field properties of double-heterostructure cavity modes. Our theory is successful on two levels: it enables accurate numerical calculations of the $Q$ factor and far-field radiation pattern of DHC modes, but more importantly, significant qualitative insight into the far-field properties of DHC modes is gained by examining the inhomogeneous driving term in the integral equation (\ref{integralEquation}). This theory has the capability to not only speed up numerical calculations but, since it provides a direct link between a cavity geometry and its far-field properties, to greatly enhance the ability to design cavities with tailored far-field properties. We have shown this by providing designs for ultrahigh $Q$ cavities whose radiation pattern has been engineered to emit predominantly in the vertical direction.

\begin{acknowledgements}
The authors thank A. Rahmani and M.J. Steel for useful discussions. This work was produced with the assistance of the Australian Research Council (ARC) under the ARC Centres of Excellence program, and was supported by an award under the Flagship Scheme of the National Computational Infrastructure of Australia, and by the Natural Sciences and Engineering Research Council of Canada (NSERC).
\end{acknowledgements}

\appendix
\section{Solving the far-field polarization integral equation}\label{app:Num}

Here we present our strategy for solving Eq.~(\ref{integralEquation}) for the polarization field $\mathbf P_1^{\rm rad}(\mathbf r)$ using a spatially discrete basis. Since $\mathbf D^a(\mathbf r)$ is constructed by superposing Bloch modes, the grid spacing is chosen such that it resolves most of the Fourier components of the Bloch modes from which it is composed. We found that a discretization of $(\Delta x, \Delta y, \Delta z)$ $=$ $(d/24, (\sqrt{3}d/2)/16, d/24)$, where $d$ is the period, was sufficient for this purpose. The Bloch modes of the even PCW band were discretized with $\Delta k \, d = 2\pi \, 0.02$, which implies a computation domain length of $50 d$ in the $x$-direction, i.e. the $x$-coordinate spans $[-25d,25d]$. In the $y$-direction our computation domain ranges between $[-16\sqrt{3}d/2,16\sqrt{3}d/2]$, while in the $z$-direction we only require points within the slab, i.e. between $[-t/2, t/2]$, with $t$ the slab thickness. Upon discretization, Eq.~(\ref{integralEquation}) becomes an inhomogeneous matrix equation
\begin{equation}
\label{matrixEq}
(\mathbf I - \mathbf E\mathbf G)\mathbf p =  \mathbf a,
\end{equation}
where $\mathbf I$ is the identity operator, $\mathbf P_1^{\rm rad}(\mathbf r) \rightarrow \mathbf p$, $\epsilon_0(\bar{\epsilon}(\mathbf r) - 1) \rightarrow \mathbf E$, $\int d \mathbf r' \, G(\mathbf r- \mathbf r';\hat{\omega}_1) \rightarrow \mathbf G$ and $\mathbf a$ contains all inhomogeneous terms. Therefore when thus represented, operators become matrices, vector fields become vectors, and a convolution with the Green tensor is represented as a matrix multiplication. In practice we compute the convolution in the $x$ and $y$ coordinates through multiplications in Fourier space. Computing the convolution directly in the $z$-direction is feasible because the slab is thin. This implies that the matrix vector product $\mathbf G \mathbf p$ computes a discrete version of Eqs.~(\ref{fieldsFromPolarization}) and (\ref{greensElectric}).

To solve Eq.~(\ref{matrixEq}) we need to compute $(\mathbf I - \mathbf E\mathbf G)^{-1} \mathbf a$. The difficulty of this inversion depends on the nature of the matrix (i.e. its symmetry properties and  sparsity, etc.) and its size. Although the matrix is sparse, from our discretization and domain size given above, the vector $\mathbf a$ has approximately $N_a \sim 10^7$ elements in each of its three components and therefore, to solve the problem, we would be required to invert a sparse matrix of size $3N_a \times 3N_a$.

Equation~(\ref{integralEquation}) has the form of the well-known discrete dipole scattering problem \cite{purcell1973scattering, draine1988discrete, draine1994discrete, flatau1997improvements}. These systems are typically too large to be solved directly and instead iterative methods are used. When the size of the problem becomes too large either the iteration does not converge, or the number of iterations becomes so large such that computations become impractical, even with the most sophisticated iterative method. No known iterative method can solve a system of equations with dimensions of $3\times 10^7$ within a reasonable computation time, so the size of the problem must be reduced.

Chamet and Rahmani \cite{chaumet2009efficient} recently tested different iterative methods for the discrete dipole problem in which a field scattering off a sphere with an electric and magnetic response. They showed that for a sphere that is discretized into $N \sim 200 000$ points, leading to $6N \times 6N$) matrices, the iterative {\it Generalized Product-type Bi-Conjugate Gradient method} (GPBiCG) \cite{zhang1997gpbi, van2003iterative} performed best in terms efficiency and robustness. Like most iterative solvers, this method works by minimizing a residual, which defines how well a solution satisfies the matrix equation. For a trial solution $\mathbf x_i$ of the matrix equation $\mathbf A \mathbf x = \mathbf b$,  the residual is defined as $\mathbf r_i = \mathbf b - \mathbf A \mathbf x_i$. The quality of the solution increases as $||\mathbf r_i|| \rightarrow 0$. Iterative methods also do not have the storage of matrix $\mathbf G$, as only matrix vector products need to be computed.

\begin{figure}[ht]
\centering\includegraphics[width=0.45\textwidth]{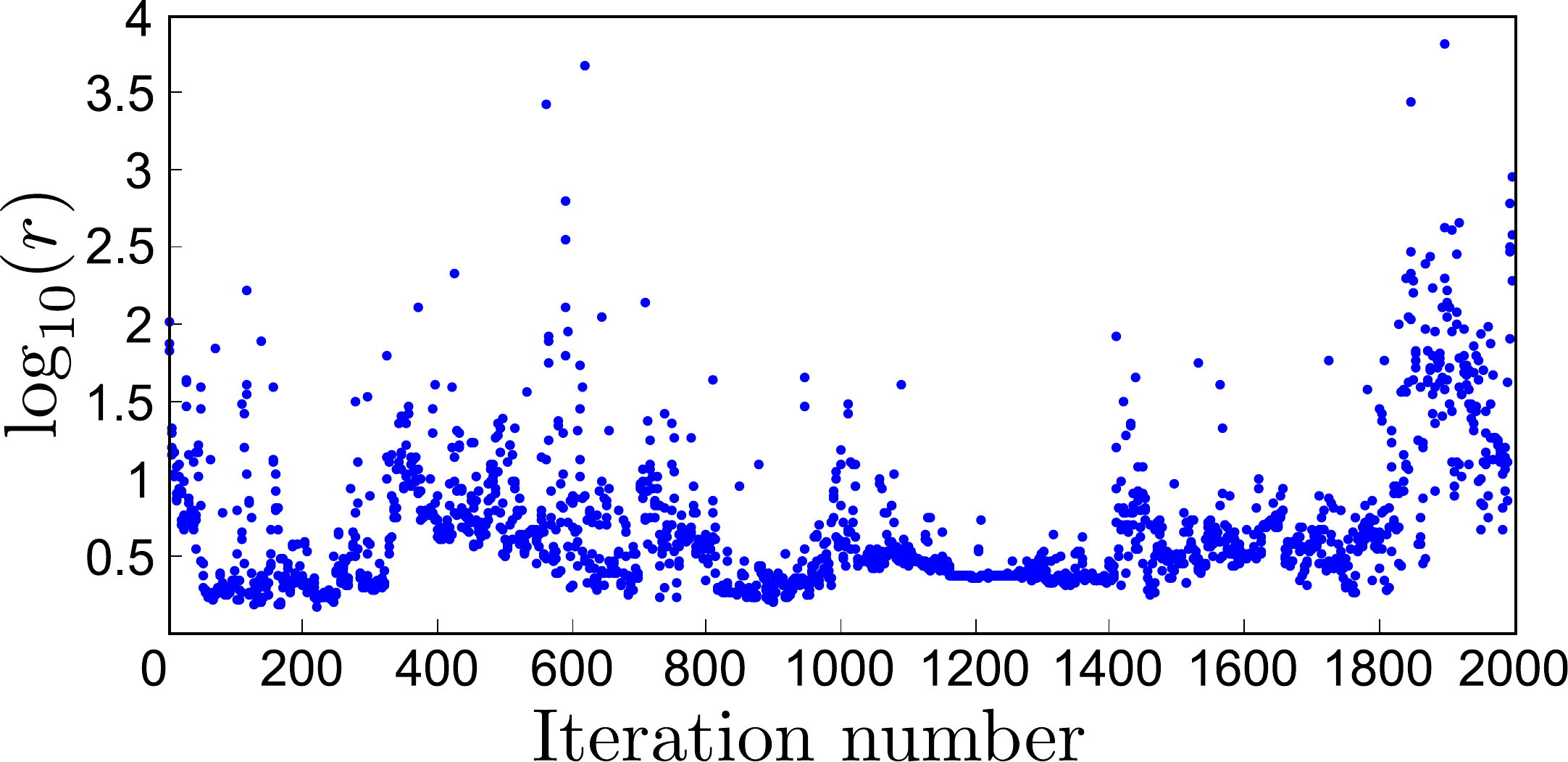}
\caption{\label{divergeIteration}(Color online) The logarithm of the residual versus iteration number for a GPBiCG algorithm applied to solve Eq.~(\ref{matrixEq}) for Cavity 2 with $\Delta n_p=0.02$ and length $L=4d$.}
\end{figure}

Since we are ultimately interested in solving for Fourier components of $\mathbf p$ inside the light cone we only require the slowly varying components of $\mathbf p$, thereby reducing the size of our problem to that in \cite{chaumet2009efficient}. Although the rapidly varying Fourier components of the field are required when finding the DHC mode $\mathbf D^a(\mathbf r)$, the radiative Fourier components are inherently slowly varying, and therefore, for this part of the problem we do not need a fine discretization in the $x-y$ directions. We reduced the discretization to $(\Delta x, \Delta y, \Delta z) = (d/4, (\sqrt{3}d/2)/4, d/24)$, where $\Delta z$ is unchanged. With this discretization each matrix vector product takes approximately $5$ seconds to compute on our MATLAB code. We note that the bulk of the computation time in the GPBiCG is taken up by the two matrix vector products in each iteration. Figure \ref{divergeIteration} shows the value of the residual versus iteration number for Cavity 2 with a cavity length $L=4d$ and $\Delta n =0.02$. This shows that even after 2000 iterations there is no sign of convergence.

Since we cannot achieve convergence we choose to solve Eq.~(\ref{matrixEq}) under an approximation: we neglect all coupling between Fourier components from inside the light cone to those outside the light cone. This means multiplying both sides of Eq.~(\ref{matrixEq}) by a Fourier filter $\hat{\mathrm F}$, that removes $k_x^2 + k_y^2>k_0^2$. Equation (\ref{matrixEq}) then becomes
\begin{equation}
\label{matrixEq2}
\hat{\mathrm F}(\mathbf I - \mathbf E\mathbf G)\mathbf p =  \hat{\mathrm F} \mathbf a.
\end{equation}
We expect solutions to Eq.~(\ref{matrixEq2}) to be approximate solutions to Eq.~(\ref{matrixEq}). This is because we have observed that multiplying a vector that only has Fourier components inside the light cone by $\mathbf E \mathbf G$, results in a vector that is still dominated by its Fourier components inside the light cone, i.e. the light cone Fourier components are weakly coupled to Fourier components outside the light cone.

\begin{figure}[ht]
\centering\includegraphics[width=0.47\textwidth]{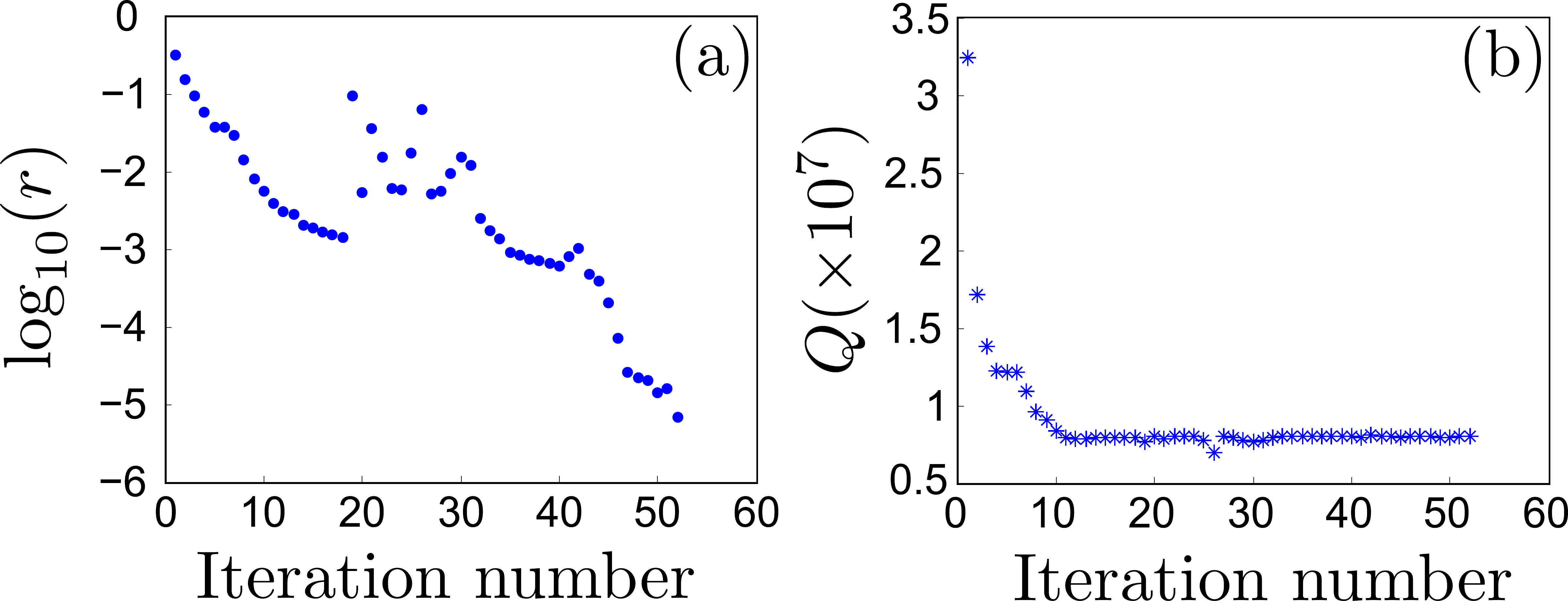}
\caption{\label{convergeIterations} (Color online) (a) The logarithm of the residual and (b) the $Q$-factor versus iteration number for a GPBiCG algorithm applied to solve Eq.~(\ref{matrixEq2}).}
\end{figure}

Typical examples of the residual and the $Q$-factor versus iteration number when solving Eq.~(\ref{matrixEq2}) using GPBiCG are shown in Figure \ref{convergeIterations}. We consider the problem to be solved when the residual is $<10^{-5}$ which, in this example, is achieved in $52$ iterations. Figure \ref{convergeIterations} shows that the calculated $Q$-factor also converges. We further test our solution by substituting it into Eq.~(\ref{matrixEq}) and checking if it satisfies this equation for the Fourier components inside the light cone--if we wish to see if $\mathbf p_t$ solves Eq.~(\ref{matrixEq}), we compute $||\mathbf p_t - (\mathbf E\mathbf G \mathbf p_t + \mathbf a)||/||\mathbf p_t||$ and check that it is of the same order as the residual (i.e $\sim 10^{-5}$). We found this to be so for all solutions presented.

\bibliography{cavRefs}

\end{document}